\documentclass[preprint,12pt]{elsarticle}

\usepackage{array}
\usepackage{booktabs}
\usepackage{tabu}
\usepackage{dcolumn}
\usepackage{amsmath}
\usepackage{amsfonts}
\usepackage{amssymb}
\usepackage{graphicx}
\usepackage{subfigure}
\usepackage{graphicx}
\usepackage{dcolumn}
\usepackage{bm}

\journal{Physics of The Dark Universe}

\begin{document}

\begin{frontmatter}



\title{Observational constraints on DBI constant-roll inflation}


\author[uok,bo]{Abolhassan Mohammadi\corref{cor1}}\ead{a.mohammadi@uok.ac.ir;abolhassanm@gmail.com}
\author[uok]{Tayeb Golanbari}\ead{t.golanbari@uok.ac.ir}
\author[uok]{Khaled Saaidi}\ead{ksaaidi@uok.ac.ir}

\cortext[cor1]{Corresponding author}
\address[uok]{Department of Physics, University of Kurdistan, Pasdaran Street, P.O. Box 66177-15175, Sanandaj, Iran.}
\address[bo]{Dipartimento di Fisica e Astronomia, Universit\`{a} di Bologna, via Irnerio 46, I-40126 Bologna, Italy. }

\begin{abstract}
Using the constant-roll approach, DBI inflationary scenario will be studied and it is seeked to compare the result with observational data. By considering the cosmological perturbations of the model, it is realized that some extra terms appear in the amplitude of scalar perturbations which indicates that there should be a modified version of scalar spectral index and tensor-to-scalar ratio. In order to compare the model with observational data, some specific functions of scalar field are assumed for the $f(\phi) $ function. For power-law and exponential functions, a constant slow-roll parameter $\epsilon$ is obtained which produces difficulties for the graceful exit from inflation. Then,  a product of linear and exponential function, and also a hyperbolic function of scalar field are selected for $f(\phi)$, that result in a $\epsilon(\phi)$ with an end for the accelerated expansion phase. Considering the scalar spectral index, amplitude of scalar perturbations, and tensor-to-scalar ratio shows that for some values of the constant $\eta=\beta$ there could be a good consistency between the model prediction and observational data. Then, based on the form of the equation of motion of scalar field a new interesting definition for the second slow-roll parameter is present and the behavior of the perturbation parameters is reconsidered. The results comes to a good agreement with observational data. Finally, the attractor behavior of these two last cases are investigated and it is determined that this feature could be satisfied.
\end{abstract}


\end{frontmatter}


\section{Introduction}\label{Intro}
The idea that the universe undergoes an extreme accelerated expansion phase in a short period of time in the first era of its evolution is well known. The idea was put forth by Alan Guth for the first time in 1981 \citep{guth} to solve the problems of hot big bang model. Since then, many inflationary scenarios have been introduced such as new inflation \citep{linde-newinflation,Albrecht}, chaotic inflation \citep{Linde-chaotic}, k-essence inflation \citep{Armendariz-Picon,Garriga,Mohammadi-d,Sheikhahmadi-b,Rezazadeh,Rezazadeh-b}, brane inflation \citep{Maartens,Golanbari}, gauge inflation \citep{Maeda,Abolhasani,Alexander,Tirandari-a,Tirandari-b}, warm inflation \citep{Berera-a,Berera-b,Taylor,Hall,Gil,Sayar,Akhtari} and so on, however a huge class of this inflationary scenarios are based on the idea of slow-rolling, proposed by A. Linde \citep{Linde-chaotic}, stating that the inflaton (the scalar field which derives inflation) slowly rolls down from the tope of its potential. So far the scenario of inflation has received a strong support from WMAP and Planck observational data \citep{WMAP,Planck2013,Planck2015,Planck2018}.  \\
Inflation usually is driven by a scalar field which in the simplest case has a potential and a canonical kinetic term. This scenario stands on slow-roll approximation so that the kinetic part is negligible in comparison to the potential part of the scalar field energy density. Then, the scalar field slowly rolls down from the top of its potential toward the minimum of it. k-inflation models are a generalized models of canonical scalar field which includes a non-canonical kinetic term. The model first introduced in \citep{Armendariz-Picon} and its cosmological perturbations was studied in \citep{Garriga}. k-essence inflation is a big class of inflation in which many inflationary models could be classified in this class such as Tachyon inflation \citep{Fairbairn,Mukohyama,Feinstein,Padmanabhan}, DBI inflation \citep{Spalinski,Bessada,Bruck,Weller,Nazavari,Amani,Rasouli} so that many works could be found in this topic \citep{Fairbairn,Mukohyama,Feinstein,Padmanabhan,Spalinski,Bessada,Bruck,Weller,Nazavari,Amani,Rasouli,Devi,Li,Peng}. In contrast to the canonical scalar field where the scalar perturbations propagate with speed of light, the scalar perturbations in k-essence inflation travel with sound speed which in general could changes \citep{Bessada,Weller,Becker,Cai}.  \\
Investigating inflationary scenario in string theory, as a theory providing a consistent formulation of quantization of gravity involving extra dimension, have received a huge interest where inflaton might be an open string \citep{Spalinski,Bessada}. This reliable theory might be our chance for understanding the fundamental characters and concepts of inflation which are missing in the standard picture \citep{Bessada}. By compacting the extra dimensions, string theory is able to anticipate a wide range of scalar fields which turns to some phenomenologically applicable inflation models such as DBI model \cite{Bessada}. In this case, the D-brane inflation, including a non-standard kinetic term, is clarified as moving of the D-brane through higher dimensions \citep{Spalinski,Bessada}. Besides the non-standard kinetic term and scalar field potential, the DBI effective action includes another function of scalar field, $f(\phi)$, which contains information about the geometry of the compact manifold traversed by the D-brane \citep{Spalinski}. An interesting feature of DBI model is that the sound speed could varies between zero and one. Then the field perturbations propagate at speed less than the speed of light which leads to this consequence that the Fourier modes freeze in at sound horizon \citep{Bruck}. The sound speed is given as inverse of the $\gamma$ parameter which plays the same role of Lorentz factor in special relativity \citep{Bruck}. \\
Slow-roll approximations are usually satisfied by almost flat part of the potential of scalar field. However, when the potential is exactly flat an strange situation occurs. From the scalar field equation of motion it could be concluded that the second slow-roll parameter becomes $\eta=-3$ that clearly breaks down the slow-roll approximation which states that the slow-roll parameters during the inflation should be smaller than unity. The situation was first studied in \citep{Kinney} where a flat potential was taken into account. The non-Gaussianity of the model was considered in \citep{Namjoo}, and they found that it is not necessary small. In \citep{Martin}, the situation was considered in more general case and the second slow-roll parameter was taken as a constant, and an approximate solution was obtained. Considering the cosmological perturbations comes to an amplitude of scalar perturbations which even vary on superhorizon scale. Taking a constant second slow-roll parameter and applying the Hamilton-Jacobi formalism \citep{Salopek,Liddle,Kinney-b,Guo-b,Aghamohammadi,Saaidi,Sheikhahmadi,Mohammadi-c} lead to an exact solution \citep{Motohashi}. Also it was realized that for specific choices of $\eta$ the amplitude of scalar perturbations could be frozen on superhorizon scale. The name "constant-roll" was first used in \citep{Motohashi}. The constant-roll inflationary scenario also has been investigated in modified gravity \citep{Motohashi-b,Awad,Anguelova,Gao,Odintsov,Oikonomou,Karam}. The scenario was generalized in \citep{Odintsov-b}, in which the second slow-roll parameter was taken as a function of scalar field and the scenario was named as "smooth-roll inflationary scenario" \citep{Odintsov-b,Oikonomou-b}. \\
The strong and interesting background of DBI scalar field motivates us to study the inflationary scenario with DBI scalar field as inflaton. In this order, the constant roll approach will be utilized where the second slow-roll parameter is taken as a constant that might not be small. The scenario of constant-roll DBI inflation was studied in \citep{Lahiri}, in which the work was limited to the constant sound speed. Taking $\eta=\dot{\epsilon}/H\epsilon$, analytical solutions for the Hubble parameter were obtained and some note about its consistency with observational were given. During the present work, we follow \citep{Weinberg,Motohashi,Mohammadi,Mohammadib}, and defined the second slow-roll parameter as $\eta=\varepsilon \ddot{\phi}/H\dot\phi$ (where $\varepsilon=\pm 1$) which turns to a different form of non-linear differential equation for the Hubble parameter. The constancy of the sound speed is released and working in ultra-relativistic regime is replaced \citep{Spalinski,Firouzjahia}. The main aim of the work is considering the model predictions with the latest observational data which will be performed in detail. Constancy of $\eta$ enforces us to reinvestigate the cosmological perturbations of the model so that a generalized form of the amplitude of scalar perturbations is obtained. This modification also appears in the scalar spectral index and tensor-to-scalar ratio. Comparison with data is carried out by introducing some specific function of the scalar field for $f(\phi)$ function, and the amplitude of scalar perturbations, scalar spectral index and tensor-to-scalar ratio are estimated and they are depicted in terms of the number of e-fold for various choice of the constant $\beta$. The results shows that some cases of $f(\phi)$ could be a good consonance with observational data.  \\
This paper is organized as follows: In Sec.\ref{SecII}, the general evolution equations of the model are presented. Then by assuming the DBI scalar field as inflaton, the equation is rewritten for DBI constant-roll inflation. Applying the assumption of constant-roll formalism, a non-linear differential equation is derived which gaining an analytical solution comes to difficulties. Then the work is limited to the ultra-relativistic regime. The cosmological perturbation of the model is studied in Sec.\ref{SecIII} where a generalized form of amplitude of scalar perturbations is obtained that only for specific choice of the constant $\eta=\beta$ is scale invariant. This modifications also appear in the scalar spectral index and tensor-to-scalar ratio so that with respect to the slow-roll inflationary scenario there are some modified terms. Computing the results of the model and comparing them with observational data for some choices of the $f(\phi)$ function are performed in Sec.\ref{SecIV} and it is realized that the model have a good agreement with observational data. In Sec.\ref{AnotherEta}, by comparing the equations of motion of scalar fiend in standard model and DBI model, a new interesting definition for $\eta$ is provided which gives a different differential equation for the Hubble parameter. Also, The perturbations parameters are reinvestigated and the results shall compared with observational data. The attractor behavior of the solution of the model is studied in Sec.\ref{Attractor}. A summary of the work also will present in Sec.\ref{conslusion}.

\section{DBI model}\label{SecII}
In this section, we are about to introduce the main equations of motion of the model. The DBI model could be categorized as a subclass of a more general one addressed as k-essence whose action is indicated by
\begin{equation}\label{action}
S = \int d^4x \sqrt{-g} \left( {1 \over 2}\; R + P(\phi,X) \right),
\end{equation}
where $g$ is the determinant of the metric tensor $g_{\mu \nu}$, R is the Ricci scalar, the inflaton field is denoted by $\phi$, and $X$ is defined as $X= g^{\mu\nu} \partial_\mu \phi \partial^\nu \phi$. The term $P(\phi,X)$ is the inflaton Lagrangian which is in general a function of $\phi$ and $X$. For DBI model, $P(\phi,X)$ is given as \citep{Silverstein,Copeland}
\begin{equation}\label{p}
P(\phi,X)= -f^{-1}(\phi) \left[ \sqrt{1-2f(\phi)\; X} - 1  \right] - V(\phi),
\end{equation}
so that $V(\phi)$ is the potential of scalar field and $f(\phi)$ is the inverse brane tension that is expressed as a function of scalar field $\phi$. \\
Obtaining the field equation of the model by taking variation of above action with respect to the metric, and applying the FLRW spatially flat metric, $ds^2 = dt^2 - a^2 \delta_{ij} dx^i dy^j$, the Friedmann equations are
\begin{equation}\label{friedmann}
  3 H^2 = \rho, \qquad  2\dot{H} = \rho + p,
\end{equation}
in which $\rho$ and $p$ are respectively the energy density and pressure DBI scalar field which are read by
\begin{eqnarray}\label{rhoandp}
  \rho &=& {\gamma -1 \over f} + V(\phi), \\
  p &=& {\gamma -1 \over f \gamma} - V(\phi),
\end{eqnarray}
and the parameter $\gamma$ is defined as
\begin{equation}\label{gamma}
\gamma=1 / \sqrt{1-f(\phi) \; \dot{\phi}^2},
\end{equation}
known as Lorentz factor \citep{Bruck,Weller}. The sound speed, which express the propagation speed of perturbations of scalar field through the homogeneous and isotropic background specetime, is obtained as \citep{Spalinski,Bessada,Bruck,Weller}
\begin{equation}\label{cs}
c_s = \sqrt{dp / d\rho} = {1 \over \gamma}.
\end{equation}
Applying the Hamilton-Jacobi approach where the scalar field is assumed as clock, the Hubble parameter is taken as a function of scalar field. Then using Eqs.(\ref{friedmann}) and (\ref{rhoandp}), the potential f the scalar field is obtained as
\begin{equation}\label{potential}
V(\phi) = 3H^2(\phi) - {\gamma - 1 \over f(\phi)}
\end{equation}
which is known as the Hamilton-Jacobi equation. From Eqs.(\ref{friedmann}), (\ref{rhoandp}) and (gamma), the time derivative of the scalar field is extracted as
\begin{equation}\label{phidot}
 \dot{\phi}^2 = {4H'^2(\phi) \over 1 + 4 f(\phi) H'^2(\phi)}
\end{equation}
Then, by substituting this in the definition of $\gamma$, the coefficient is rewritten as
\begin{equation}\label{gamma02}
\gamma = \sqrt{1 + 4f(\phi) H'^2(\phi)}
\end{equation}

\subsection{DBI constant-roll inflation}
Inflation is known as a short period of accelerated expansion while the Universe undergoes an extreme expansion. The acceleration equation of the universe is usually given by $\ddot{a}/a = H^2(1-\epsilon)$ where $\epsilon=-\dot{H}/H^2$. Therefore, the condition $\epsilon < 1$ implies an acceleration phase for the Universe, and the equality $\epsilon = 1$ is usually taken as the end of inflation. This parameter is known as the first slow-roll parameter. Following \citep{Weinberg,Motohashi,Mohammadi,Mohammadib}, the second slow-roll parameter is defined as the rate of variation of $\dot{\phi}$ during a Hubble time
\begin{equation}\label{eta}
\eta = {\varepsilon \ddot{\phi} \over H \dot{\phi}}
\end{equation}
where $\varepsilon=\pm 1$, mostly because there is an arbitrariness in putting the negative sign. In constant-roll approach of studying inflationary scenario, the second slow-roll parameter is taken as a constant. Then, a second order non-linear differential equation for the Hubble parameter is found out as
\begin{equation}\label{hubbleequation}
H''(\phi) - 2f'(\phi) H'^3(\phi) - {\beta \over 2}\; H \; \left( 1 + 4f(\phi) H'^2(\phi) \right)^{3/2} \; = 0.
\end{equation}
where $\beta$ is the constant, i.e. $\eta=\beta$. Solving above equation and finding an analytical solution seems unlikely. So, the equation will be investigated in the ultra-relativistic regime where the quantity $\gamma$ is large \citep{Spalinski,Firouzjahia}. Then we have
\begin{equation}\label{gamma-urr}
 \gamma \simeq 2H'(\phi) \sqrt{f(\phi)}, \qquad {\rm and} \qquad \dot{\phi} \simeq {-1 \over \sqrt{f(\phi)}}.
\end{equation}
Using the assumption of constant $\eta$ and Eq.(\ref{eta}), one can obtain the Hubble parameter versus of $f(\phi)$ as
\begin{equation}\label{hubble}
H(\phi) = {\varepsilon \over 2 \beta} \; {f'(\phi) \over  f^{3/2}(\phi)}.
\end{equation}
Successfully solving the standard big bang theory problems requires enough expansion for the universe in its earliest time that is assumed to be explained by inflation. The universe expansion is measured by the number of e-fold, given by
\begin{equation}\label{efold}
N = \int_{t_\star}^{t_e} H dt = {-\varepsilon \over 2\beta} \int_{\phi_\star}^{\phi_e} {f'(\phi) \over f(\phi)} \; d\phi ,
\end{equation}
where the subscribe "$e$" indicates the quantity at the end of inflation, and "$\star$" denotes the time of horizon crossing.

\section{Cosmological perturbation}\label{SecIII}
Cosmological perturbations are an interesting prediction of inflationary scenario, which generally are divided to three types as: scalar, vector and tensor perturbations. Up to the first order of perturbations parameters, these types of perturbation are evolved independently. In this section, the cosmological perturbations of the model will be considered by imposing this assumption that the second slow-roll parameter is constant and might not be small. Following \citep{Garriga} and applying a small inhomogeneous perturbation for the scalar field $\phi(t,\mathbf{x})=\phi_0(t) + \delta\phi(t,\mathbf{x})$, the metric tensor will be disturbed as well, where in Newtonian gauge it is read by
\begin{equation}\label{perturbedmetric}
 ds^2 = \big( 1 + 2\Phi(t,\mathbf{x}) \big) dt^2 - a^2(t) \big( 1 - 2\Phi(t,\mathbf{x}) \big) \gamma_{ij} dx^i dx^j ,
\end{equation}
Note that it is presumed that the perturbed energy-momentum tensor is diagonal, $\delta T^i_{\ j} \propto \delta^i_{\ i}$. Substituting the above metric in the field equations, the $(0,0)$ and $(0,i)$ component of perturbed equations are obtained as \citep{Garriga}
\begin{eqnarray}\label{pertubedequations}
\dot{\xi} & = & {a (\rho + p) \over H^2} \; \zeta, \nonumber \\
\dot{\zeta} & = & {c_s^2 H^2 \over a^3 (\rho + p)} \nabla^2 \xi,
\end{eqnarray}
where the variables $\xi$ and $\zeta$ are defined as
\begin{eqnarray*}
  \xi &\equiv& {a \over 4\pi G H} \; \Phi, \\
  \zeta &\equiv& {4\pi G H \over a}\; \xi + H {\delta \phi \over \dot{\phi}} = \Phi + H {\delta \phi \over \dot{\phi}}.
\end{eqnarray*}
The corresponding action for the equations (\ref{pertubedequations}) is
\begin{equation}\label{zetaaction}
  S = {1 \over 2} \int z^2 \left( \zeta'^2 + c_s^2 \zeta (\nabla\zeta)^2 \right) d\tau d^3\mathbf{x},
\end{equation}
here the prime denotes derivative with respect to the conformal time $\tau$, and $z$ is defined as $z=a\sqrt{\rho+p}/c_s H$. By defining a canonical quantization variable $v=z\zeta$, the above action is given as follows \citep{Garriga}
\begin{equation}\label{vaction}
  S = {1 \over 2} \int  \left( v'^2 + c_s^2 v (\nabla v)^2 + {z'' \over z} v \right) d\tau d^3\mathbf{x}.
\end{equation}
Then, the dynamical equation for the variable $v$ is obtained as
\begin{equation}\label{vxequation}
v''(t,\mathbf{x}) - c_s^2 \nabla^2 v(t,\mathbf{x}) - {z'' \over z}\; v(t,\mathbf{x}) =0,
\end{equation}
and by using the Fourior mode, one has
\begin{equation}\label{vkequation}
v''_k(\tau) + \Big( c_s^2 k^2 - {z'' \over z}\; \Big) \; v_k(\tau) =0.
\end{equation}
The term $z''/z$ up to the first order of the slow-roll parameters $\epsilon$ and $s$ could be expressed as follows
\begin{eqnarray}
z &=& {a\sqrt{\rho+p} \over c_s H} = {a\sqrt{\gamma \dot{\phi}^2} \over c_s H}, \hspace{4cm} \\
z' &=& z \big( aH \big) \Big( 1 + \epsilon + \varepsilon \eta - {3 \over 2} s \Big), \\
z'' &=& z \big( aH \big)^2 \Big( 2 + 6\epsilon + 3\varepsilon \eta - 3s + 9\varepsilon \epsilon \eta - 3\varepsilon s \eta \nonumber \\
    & & \hspace{4cm} + \eta^2 + 2\epsilon \eta^2 \Big),
\end{eqnarray}
where the slow-roll parameter $s$ is given as
\begin{equation}\label{s}
s = {\dot{c}_s \over H c_s}.
\end{equation}
Utilizing the variable changes $x = -c_sk\tau$ and $v_k = \sqrt{-\tau} \; f_k(\tau)$ the Eq.(\ref{vkequation}) could be transformed to the Bessel differential equation
\begin{equation}\label{besseldifferentialequation}
{d^2 f_k \over dx^2} + {1 \over x}{d f_k \over dx} + \Big( 1 - {\nu^2 \over x^2} \Big)f_k=0,
\end{equation}
in which we have used $a H = {-(1+\epsilon) \over \tau}$ and the parameter $\nu$ is defined by ${z'' \over z} ={\nu^2 - {1 \over 4} \over \tau^2}$. The general solution is a combination of first and second type of Hankel function
\begin{equation}\label{hankel}
f_k(\tau) = \alpha_1(k) H_\nu^{(1)}(-k\tau) + \alpha_2(k) H_\nu^{(2)}(-k\tau).
\end{equation}
where $\alpha_1(k)$ and $\alpha_2(k)$ are constant that could be determined by considering the asymptotical behavior
of the equation. \\
In subhorizon scale, where $c_s^2k^2 \gg a^2 H^2$ or in another word $c_s^2k^2 \gg z''/z$, the differential equation (\ref{vkequation}), could be stated as
\begin{equation}\label{vksubhorizon}
v''_k(\tau) + c_s^2 k^2 \; v_k(\tau) =0,
\end{equation}
and the solution is obtained as
\begin{equation}\label{vksubsolution}
 v_k(\tau) = {1 \over \sqrt{2 c_s k}} \; e^{-ic_s k \tau}.
\end{equation}
It could be concluded that the general solution (\ref{hankel}) should return to the subhorizon solution (\ref{vksubsolution}) for scale $c_s^2k^2 \gg a^2 H^2$. Then, by studying the asymptotical behavior of Hankel function, it is realized that the constant $\alpha_2(k)$ should be eliminated, $\alpha_2(k)=0$, and for $\alpha_1(k)$ there is $\alpha_1(k)= {\sqrt{\pi} \over 2} \; e^{{\pi \over 2}\; (\nu + {1 \over 2})}$. On the other hand, at superhorizon limit, the solution is read as
\begin{equation}\label{vksupersolution}
\lim_{-k\tau \rightarrow 0} = {2^{\nu - {3 \over 2}} \Gamma(\nu) \over \sqrt{2 c_s k} \; \Gamma(3/2)}\; e^{{\pi \over 2}\; (\nu - {1 \over 2})} \; \big( -c_s k \tau \big)^{{1 \over 2}-\nu}.
\end{equation}
The amplitude of scalar perturbation is defined as
\begin{equation}\label{psdefinition}
\mathcal{P}_s^{1/2} = \sqrt{k^3 \over 2\pi^2} \; \Big|\zeta\Big| = \sqrt{k^3 \over 2\pi^2} \; \Big|{v_k \over z}\Big|.
\end{equation}
Then, one could arrive at
\begin{eqnarray}\label{pssuper}
\mathcal{P}_s \Big|_{superhorizon} & = &  A_s \; \left( {c_s k \over a H} \right)^{3-2\nu}  \\
   & = & {1 \over 8\pi^2} \left( {2^{\nu - {3 \over 2}} \Gamma(\nu) \over  \; \Gamma(3/2)} \right)^2 \;
    {H^2 \over c_s \epsilon } \; \left( {c_s k \over a H} \right)^{3-2\nu}. \nonumber
\end{eqnarray}
The scalar spectral index, as an important observational parameter is defined through the relation of amplitude of scalar perturbation so that $\mathcal{P}_s = A_s \left( {c_s k \over a H} \right)^{n_s-1}$, in which $A_s$ is the amplitude of scalar perturbation at horizon crossing. Then, there is
\begin{equation}\label{ns}
ns-1 = 3- 2\nu,
\end{equation}
and
\begin{equation}\label{nu2}
\nu^2= {9 \over 4} + 6\epsilon + 3\varepsilon \eta - 3s + 9\varepsilon \epsilon\eta - 3\varepsilon s \eta + \eta^2 + 2\epsilon \eta^2.
\end{equation}

Since energy-momentum tensor has no contribution in the tensor perturbations equations, and only the metric perturbations, $\delta_{ij} = a^2 h_{ij}$, play the roles, one has the following well-known evolution equations for the mode function of the tensor perturbations \cite{Motohashi}
\begin{equation}\label{tensorperturbationequation}
u''_{k,\lambda} + \left( k^2 - {a'' \over a} \right) \; u_{k,\lambda}=0
\end{equation}
in which $u_{k,\lambda} = a h_{k,\lambda} / 2$, and $\lambda=+,\times$ indicates two polarization modes of the gravitational waves. The term ${a'' \over a}$ is given as ${a'' \over a}= (a H)^2 (2-\epsilon)$ \cite{Motohashi}, so it is realized that only the first slow-roll parameters appears and the amplitude of the tensor perturbations is expected to be same as slow-roll inflationary scenario as $\mathcal{P}_t = 2H^2 / \pi^2$. \\
Tensor perturbations are measured indirectly through the parameter $r$ which is stated as the ratio of tensor perturbations to the scalar perturbations, $r=\mathcal{P_t} / \mathcal{P}_s$, so that at horizon crossing one can obtain
\begin{equation}\label{r}
r= 16 \; \left( { \Gamma(3/2) \over  2^{\nu - {3 \over 2}} \; \Gamma(\nu)} \right)^2 c_s \epsilon .
\end{equation}

\section{Observational Constraint}\label{SecIV}
After considering the main evolution equations of the model, and discussing the cosmological perturbations of the model, we are going to compare the model prediction with observational data. In this regards, it is required to specify an specific function of scalar field for $f(\phi)$. Then, in the following lines, four typical examples of $f(\phi)$ will be introduced and for each one the consequences shall be discussed.

\subsection{Power-law function}
A power-law function of scalar field is taken as the first case, in which $f(\phi) = f_0 \phi^n$ where $f_0$ and $n$ are real constants. By substituting it in Eq.(\ref{hubble}), the Hubble parameter is found in terms of the scalar field as
\begin{equation}\label{hubble01}
H(\phi) = {n \varepsilon \over 2\beta \sqrt{f_0}} \; {1 \over \phi^{{n \over 2}+1}}.
\end{equation}
Then, the first slow-roll parameter could be obtained using Eq.(\ref{gamma-urr}) and (\ref{hubble01}) as
\begin{equation}\label{epsilon01}
\epsilon = {-\dot{H} \over H^2} = {\dot\phi H' \over H^2} = \varepsilon \beta \; \Big( 1 + {1 \over 2n} \Big)
\end{equation}
which is a constant. This constant might be smaller than unity and give a positive accelerated expansion phase however, graceful exit from inflation encounters difficulties. Due to this fact, one can see that the exact power-law choice for $f(\phi)$ is not an appropriate choice.

\subsection{Exponential function}
As a second case, an exponential function of the scalar field is selected for $f(\phi)$, such as $f(\phi) = f_0 e^{\lambda\phi}$ where $f_0$ and $\lambda$ are real constants. The Hubble parameter is achieved from Eq.(\ref{hubble})
\begin{equation}\label{hubble02}
H(\phi) = {\varepsilon \lambda \over 2\beta \sqrt{f_0}} \; e^{-\lambda \phi \over 2}.
\end{equation}
From Eq.(\ref{gamma-urr}), the time derivative of the scalar field is acquired as
$\dot\phi = -e^{-\lambda \phi \over 2}/\sqrt{f_0}$. Then, utilizing $\dot{H}= \dot\phi H'$, the first slow-roll parameter is given by
\begin{equation}\label{epsilon02}
\epsilon = - 2 \varepsilon \beta .
\end{equation}
A proper choice of $\beta$ leads to an accelerated expansion phase, but there is the same problem that we have for the previous case and exiting from inflation seems unlikely.

\subsection{Combination function}
\begin{figure}[ht]
  \centering
  \subfigure[]{\includegraphics[width=7cm]{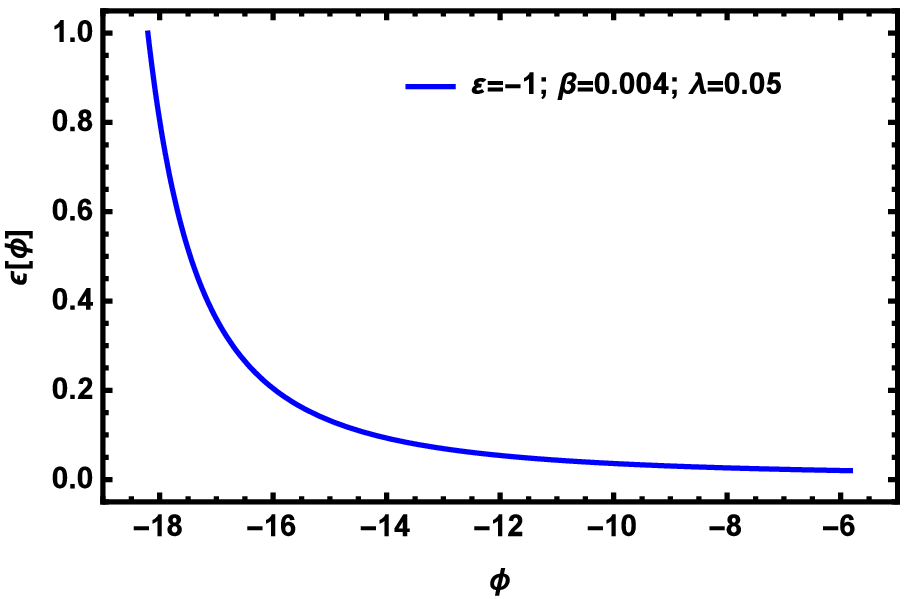}}
  \subfigure[]{\includegraphics[width=7cm]{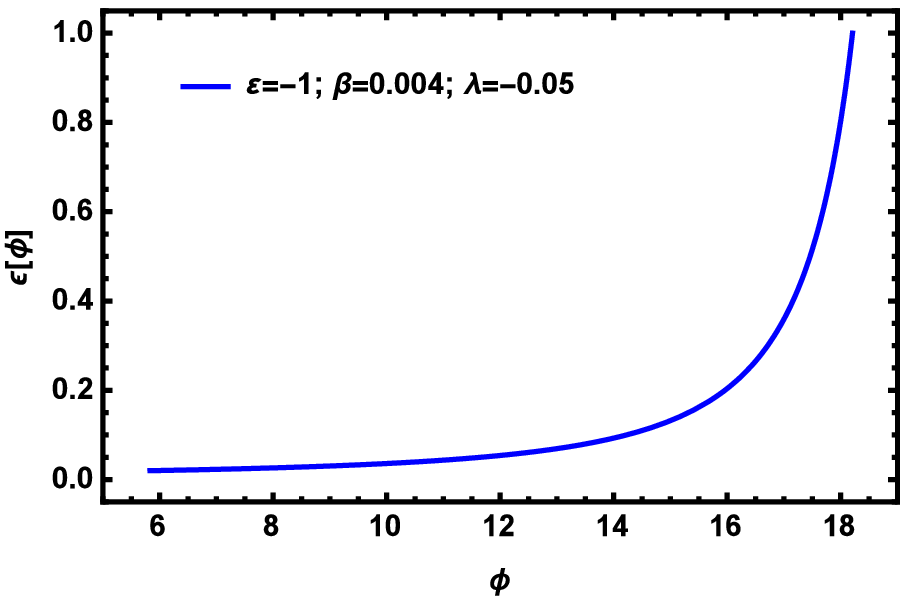}}
  \subfigure[]{\includegraphics[width=7cm]{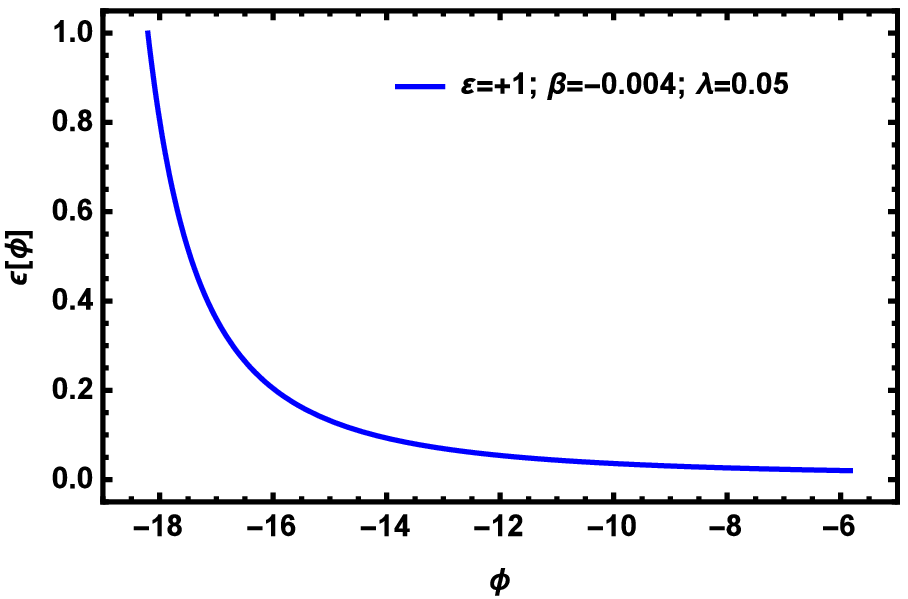}}
  \subfigure[]{\includegraphics[width=7cm]{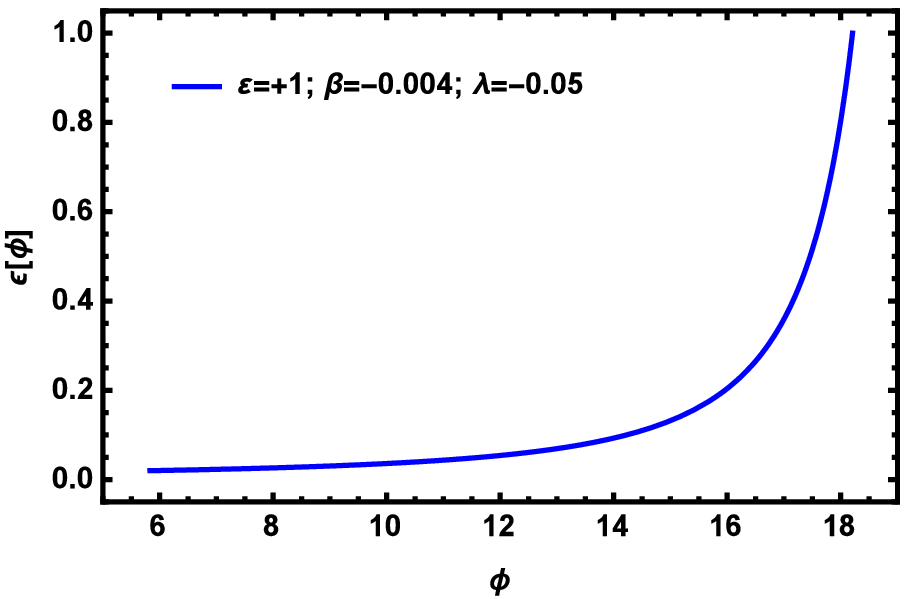}}
  \caption{The behavior of the slow-roll parameter $\epsilon$ versus the scalar field during the inflationary times for different values of the model parameters.}\label{figepsilon03}
\end{figure}

\begin{figure}[ht]
  \centering
  \subfigure[]{\includegraphics[width=7cm]{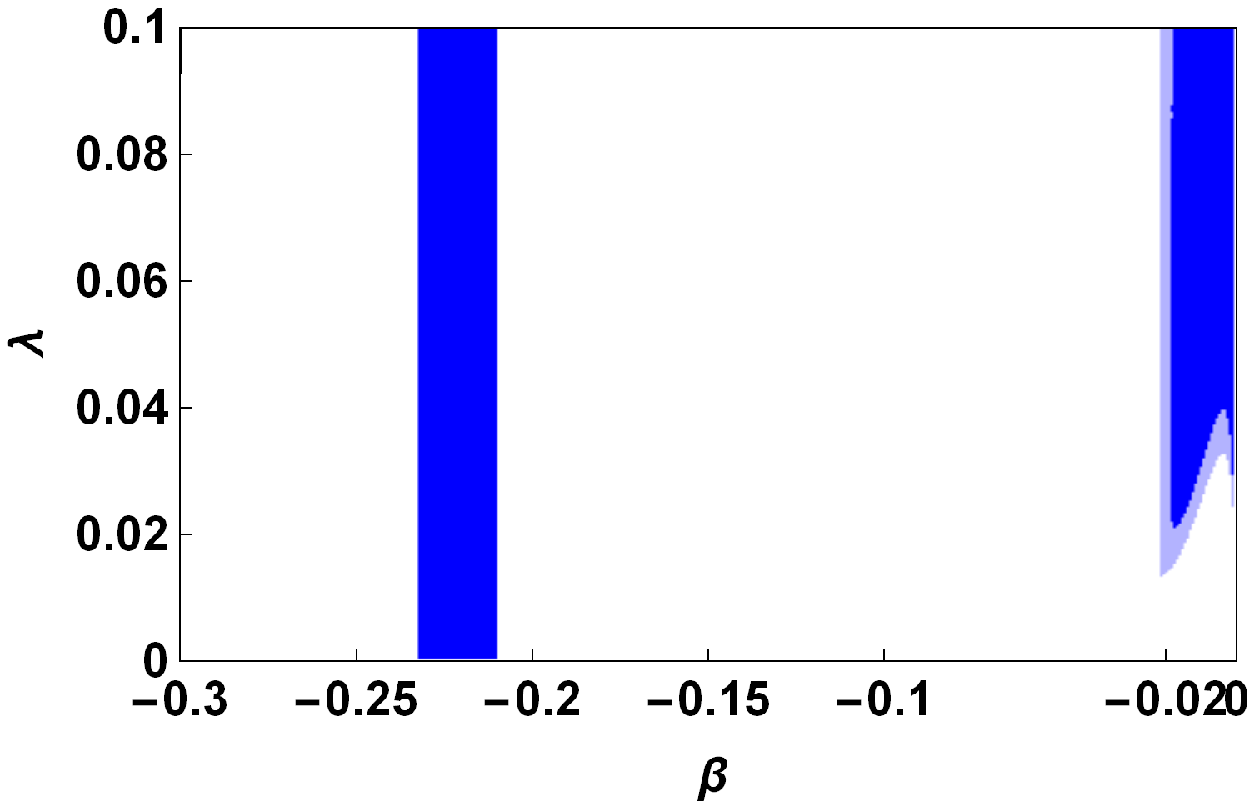}}
  \subfigure[]{\includegraphics[width=7cm]{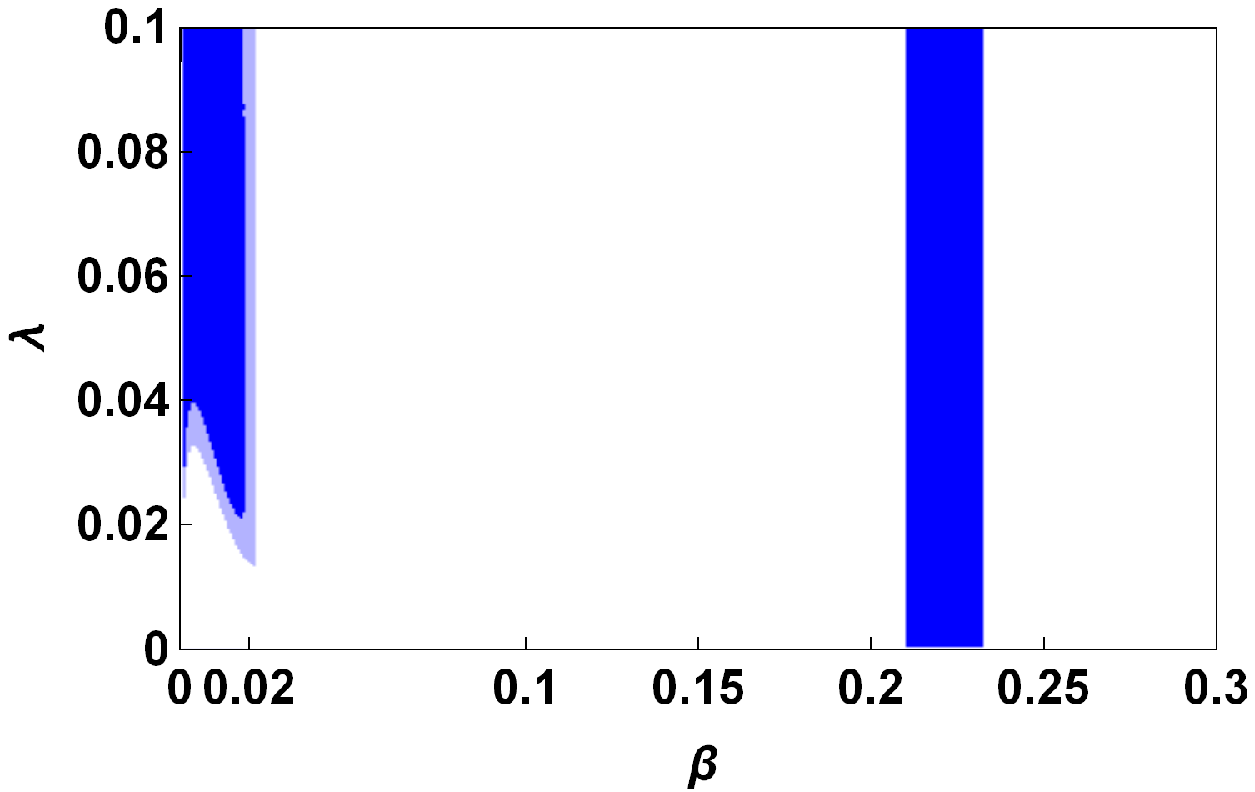}}
  \caption{The valid range of the constants $\beta$ and $\lambda$ to have a consistent result with data for the scalar spectral index and tensor-to-scalar ratio; where in a) $\varepsilon=+1$ and b) $\varepsilon=-1$.}\label{BetaLambda03}
\end{figure}

\begin{figure}
\centering
\includegraphics[width=7cm]{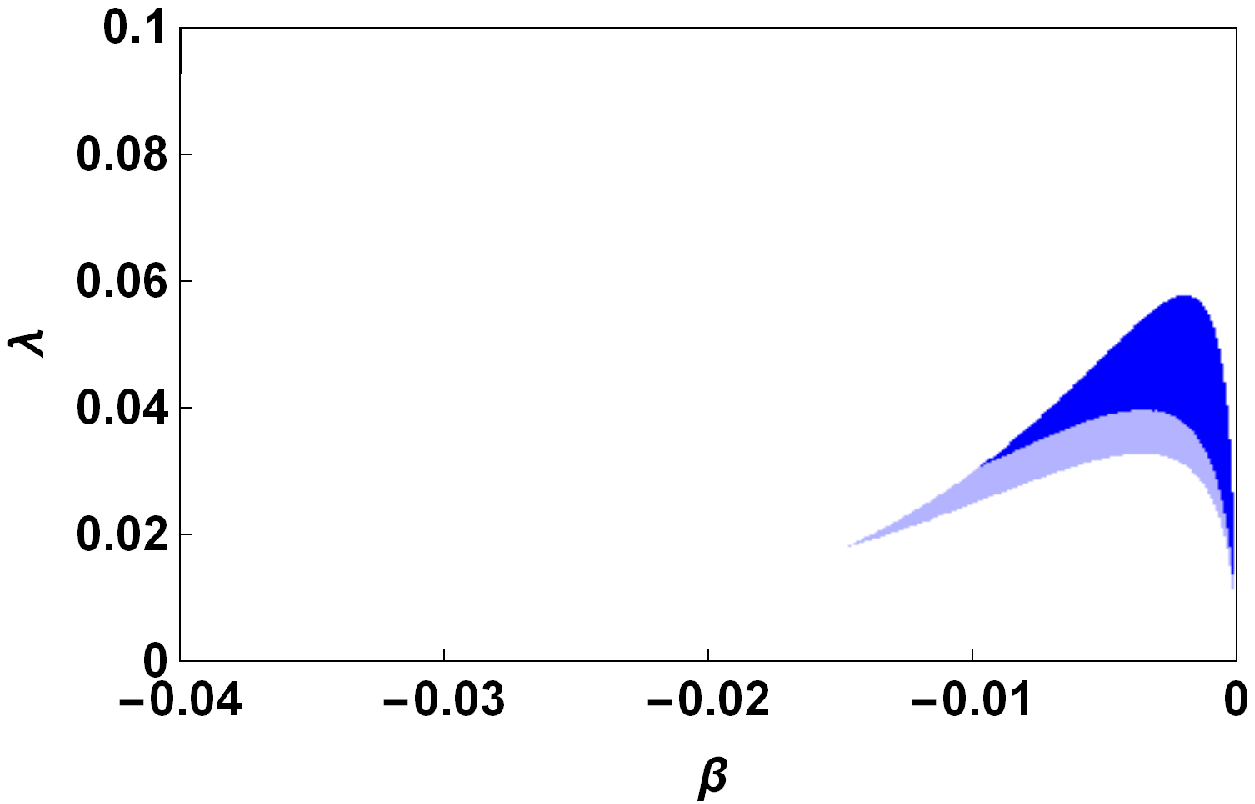}
\caption{Parametric space of $\beta$ and $\lambda$ to give the acceptable values for $n_s$ and $r$, and sound speed $0.1<c_s<1$}\label{csbetalambda03}
\end{figure}

In this case, we combine two previous choice and take the function $f(\phi)$ as $f(\phi)=f_0 \phi e^{\lambda\phi}$. From Eq.(\ref{hubble}), the Hubble parameter is derived in terms of the scalar field as
\begin{equation}\label{hubble03}
H(\phi) = {\varepsilon \over 2\beta} \; {(1+\lambda\phi) e^{-\lambda\phi/2} \over \phi \sqrt{f_0 \phi}}.
\end{equation}
Using above relation and applying the time derivative of scalar field (\ref{gamma-urr}), the first slow-roll parameter for this case is given by
\begin{equation}\label{epsilon03}
\epsilon = 2\varepsilon \beta \left[ {\lambda \phi (2 + \lambda \phi) \over (1 + \lambda \phi)^2} - {3 \over 2} \right].
\end{equation}
which is a function of scalar field. \\
Inflation ends as $\epsilon$ reaches unity, which occurs for scalar field $\phi_e$
\begin{equation}\label{phiend03}
\lambda \phi_e = -1 \pm {\sqrt{1 - \sigma} \over 1 - \sigma}, \qquad \sigma = {\varepsilon \over 2\beta} + {3 \over 2}.
\end{equation}
Then, to have a physical answer, the condition $\sigma < 1$ should be satisfied which states that $\epsilon$ and $\beta$ must have a different sign. In order to solve the horizon and flatness problem, about $N=60-70$ number of e-fold is required. To have this amount of expansion, reading from Eq.(\ref{efold}), the scalar field at the horizon crossing is obtained as
\begin{equation}\label{phistar03}
\lambda \phi_\star = W\left[ \lambda \phi_e \; e^{\lambda \phi_e} \; e^{2\varepsilon\beta N}  \right],
\end{equation}
where $W$ indicates the Lambert function. Fig.\ref{figepsilon03} displays the behavior of the slow-roll parameter $\epsilon$ versus scalar field for different choices of $\varepsilon$, $\beta$ and $\lambda$. For all cases, $\varepsilon$ and $\beta$ have a different sign to have a physical situation. Fig.\ref{figepsilon03}a and b describe the parameter $\epsilon(\phi)$ for positive $\lambda$ indicating that $\epsilon(\phi)$ reaches one by decreasing the scalar field. It is consistent with the negative time derivative of the scalar field that we have in Eq.(\ref{gamma-urr}). However, for negative $\lambda$, the parameter $\epsilon(\phi)$ approaches one by increasing the scalar field which requires a positive $\dot\phi$; opposes to Eq.(\ref{gamma-urr}). Then, we are going to get $\lambda$ as a positive constant for the rest of this part. Therefore, for this case the inflation happens for negative values of the scalar field. \\
\noindent The other slow-roll parameter which is needed for estimating the scalar spectra index is $s$. Therefore, using Eq.(\ref{s})one could find it as
\begin{equation}\label{s01}
s = {\dot{c}_s \over H c_s} = {-\dot\phi \over H} \; \left( {H'' \over H'} + {f' \over 2f} \right),
\end{equation}
and by using the function $f(\phi)$ and the Hubble parameter (\ref{hubble03}), one arrives at
\begin{eqnarray}\label{s011}
s & = &  2\varepsilon \beta \left\{ { \lambda \phi \left[ \left( {1 \over 1 + \lambda\phi} - {1 \over 2} - {3 \over 2\lambda\phi} \right)^2 + \left( {3 \over 2\lambda^2 \phi^2} - {1 \over (1+\lambda\phi)^2} \right) \right] \over  \left( {1 \over 1 + \lambda\phi} - {1 \over 2} - {3 \over 2\lambda\phi} \right)} \right. \nonumber \\
   &  &  \qquad \qquad \qquad + {(1 + \lambda\phi) \over 2} \Biggl\}.
\end{eqnarray}
The second slow-roll parameter is $\eta=\beta$ which is a real constant. These slow-roll parameters are very important when the predictions of the model at horizon crossing are compared with observational data, which will be discussed next. \\
One of the most important observational data is $r-n_s$ diagram which determine the scalar spectral index and tensor-to-scalar ratio \citep{Planck2018}. To compare our result with observational data, we are about to extract the model constants which put the scalar spectral index and tensor-to-scalar ratio of the model in agreement with data. In this regards, the slow-roll parameters should be specified at the time of horizon crossing by using Eq.(\ref{phistar03}). After some manipulation, it is concluded that $n_s$ and $r$ only depend on the constants $\beta$ and $\lambda$. Then, utilizing the $r-n_s$ diagram of Planck, one could determine the valid range of these two parameters which are illustrated in Fig.\ref{BetaLambda03} for $\varepsilon=+1$ and $\varepsilon=-1$. There is the same results for both choices of $\varepsilon$ however with different sign for $\beta$. 
The other constant of the model is $f_0$ which could be clarify through the curvature perturbations. The latest observational data states that the amplitude of scalar perturbations should be about $\ln\left( 10^{10} A_s \right) = 3.044 \pm 0.014$ \citep{Planck2018}. Calculating the amplitude of curvature perturbation at the time of horizon crossing, the constant $f_0$ is read as
\begin{equation}\label{f03}
f_0 = \left( {2^{\nu - {3 \over 2}} \; \Gamma(\nu) \over \Gamma(3/2)} \right)^2 \; {\lambda^3 (1+\lambda\phi_\star)^2 r^{-\lambda\phi_\star} \over 4 \beta^2 (\lambda\phi_\star)^3 \left( 8 \pi^2 c_s^\star \epsilon^\star \mathcal{P}_s^\star \right) }
\end{equation}
where the values of the constants $\beta$ and $\lambda$ are given by Fig.\ref{BetaLambda03}. Therefore, all the free parameters of the model have been specified, and one could use Eq.(\ref{potential}) for the scalar field potential and Eq.(\ref{cs}) to obtain the energy scale of the inflation and sound speed (where $c_s=1/\gamma$). Table.\ref{table03} shows these values for different choices of $\lambda$ and $\beta$ for number of e-fold $N=65$.
\begin{table}
  \centering
  \caption{The estimated values of the constant $f_0$, energy scale of inflation, and sound speed for different values of $\beta$ and $\lambda$.}

\begin{tabular}{lllll}
\hline
\hline
  \qquad $\beta$ &  $\ \ \lambda$  & \ \ \qquad $f_0$  & \qquad $V^\star$ & $\ c_s^\star$\\
\hline
 $-0.0015$ & $0.05$  & $-1.60\times 10^{11}$ & $8.69\times 10^{-10}$ & $0.13$ \\
 $-0.0015$ & $0.08$  & $-1.69\times 10^{12}$ & $3.39\times 10^{-10}$ & $0.05$ \\
 $-0.004$  & $0.05$  & $-1.46\times 10^{11}$ & $1.07\times 10^{-9}$ & $0.11$ \\
 $-0.004$  & $0.08$  & $-1.54\times 10^{12}$ & $4.18\times 10^{-10}$ & $0.042$ \\
 $-0.015$  & $0.03$  & $-6.44\times 10^{11}$ & $7.90\times 10^{-10}$ & $0.034$ \\
 $-0.015$  & $0.05$  & $-8.28\times 10^{12}$ & $2.87\times 10^{-10}$ & $0.012$ \\
 $-0.216$  & $0.001$ & $-1.05\times 10^{58}$ & $4.28\times 10^{-28}$ & $2.3\times 10^{-21}$ \\
 $-0.216$  & $0.004$ & $-1.06\times 10^{61}$ & $2.68\times 10^{-29}$ & $1.4\times 10^{-22}$ \\
  \hline
\end{tabular}

\label{table03}
\end{table}
The potential of the scalar field is about $10^{-10}$ which states that the energy scale of inflation could be around $10^{-3}(M_p)$. In a general look, it is realized that by increasing $\lambda$ for every specific values of $\beta$, the constant $f_0$ increases (with a negative sign), and the potential energy and the sound speed decrease. Turing attention to the sound speed of the model, it is realized that depending to the values of $\beta$ and $\lambda$, thesound speed could be $>0.1$ or $<0.1$. To get a more optimum range for the constants $\beta$ and $\lambda$ to even get a proper sound speed, the condition $0.1<c_s<1$ could be imposed in addition to the scalar spectral index and tensor-to-scalar ratio equations which certainly is resulted in more restricted ranges for the constants. Fig.\ref{csbetalambda03} illustrates the consequence of such a procedure where it is seen that there are small ranges of $\beta$ and $\lambda$ which come to acceptable results for scalar spectral index, amplitude of curvature perturbations, tensor-to-scalar ratio, sound speed and the energy scale of inflation.

\subsection{Hyperbolic function}
\begin{figure}
  \centering
  \subfigure[]{\includegraphics[width=7cm]{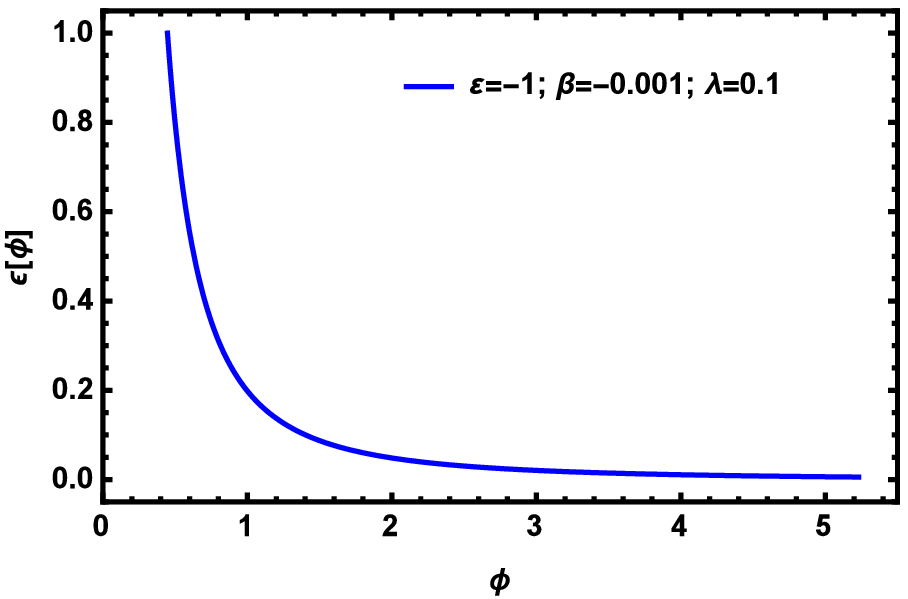}}
  \subfigure[]{\includegraphics[width=7cm]{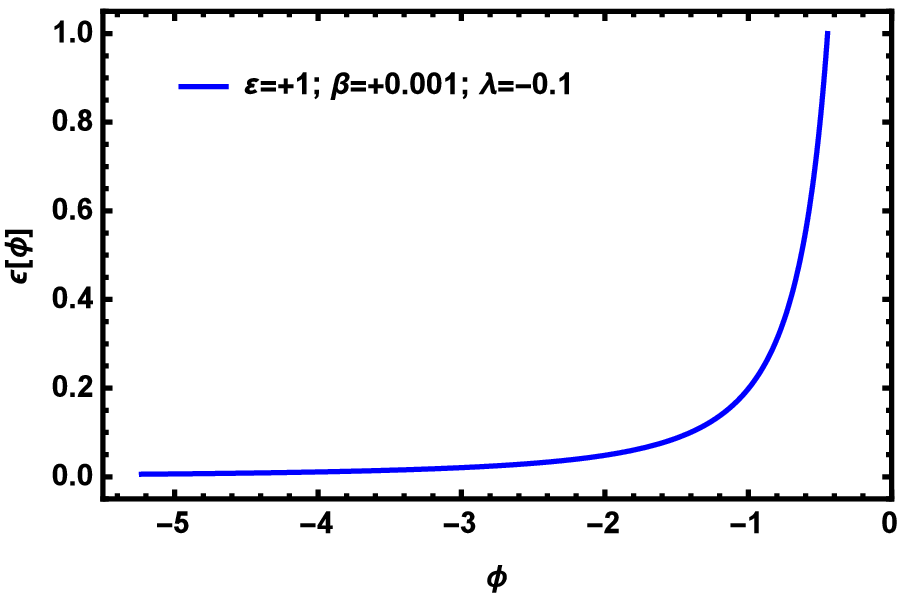}}
  \subfigure[]{\includegraphics[width=7cm]{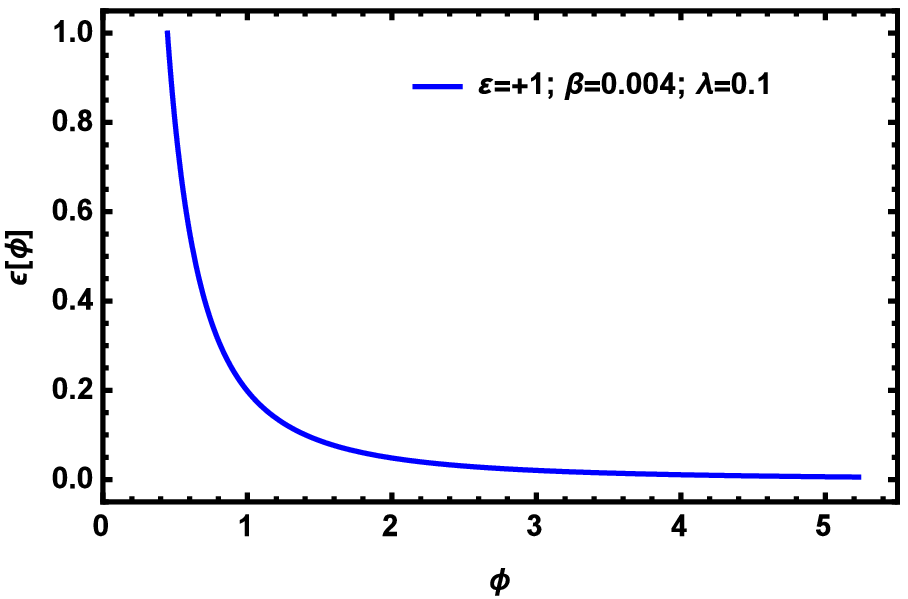}}
  \subfigure[]{\includegraphics[width=7cm]{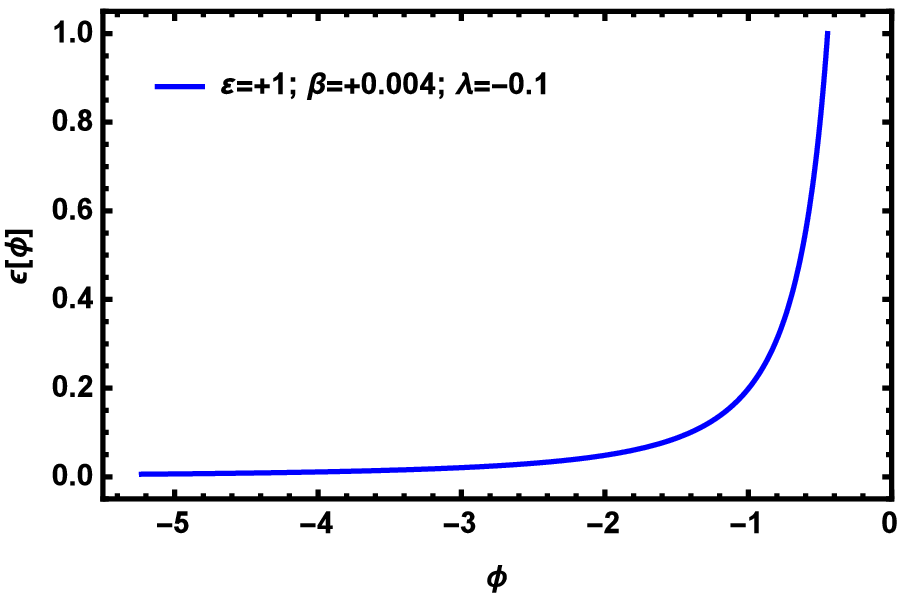}}
  \caption{The slow-roll parameter $\epsilon$ is plotted in terms of $\phi$ during inflation for different values of the model parameters.}\label{figepsilon04}
\end{figure}

\begin{figure}[ht]
  \centering
  \subfigure[]{\includegraphics[width=7cm]{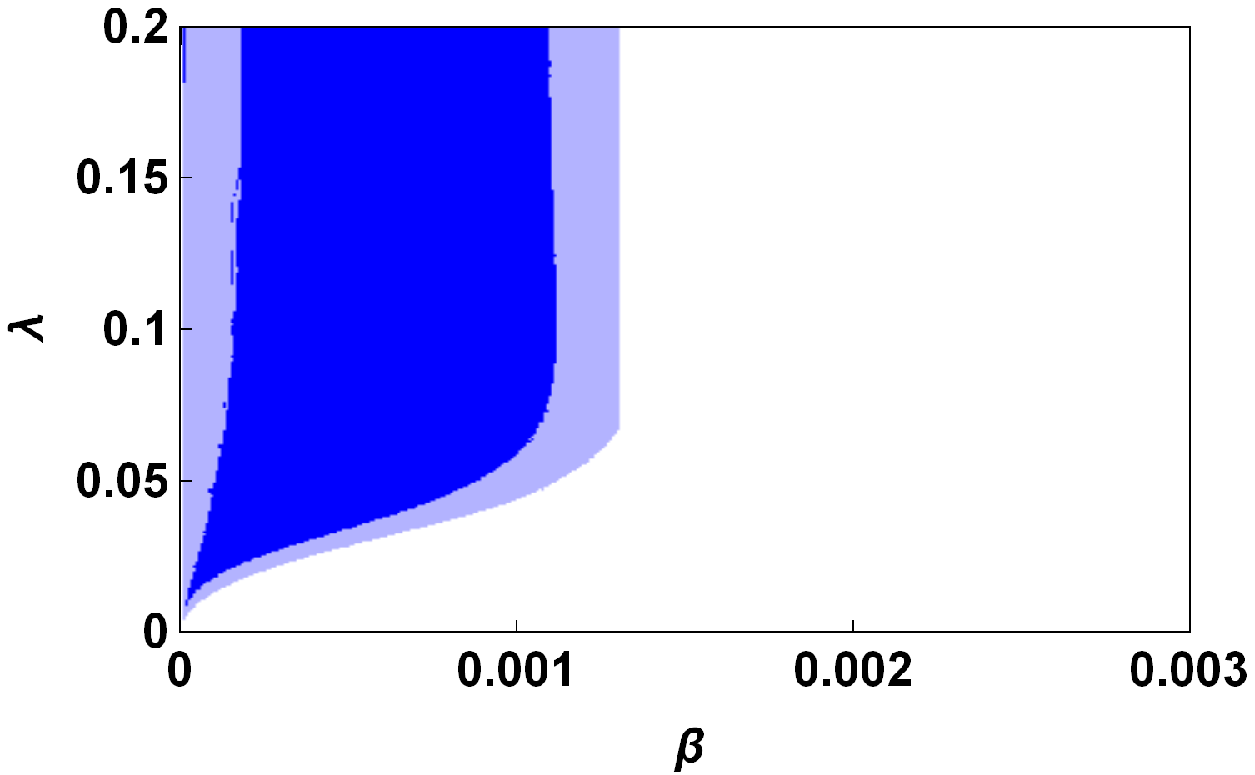}}
  \subfigure[]{\includegraphics[width=7cm]{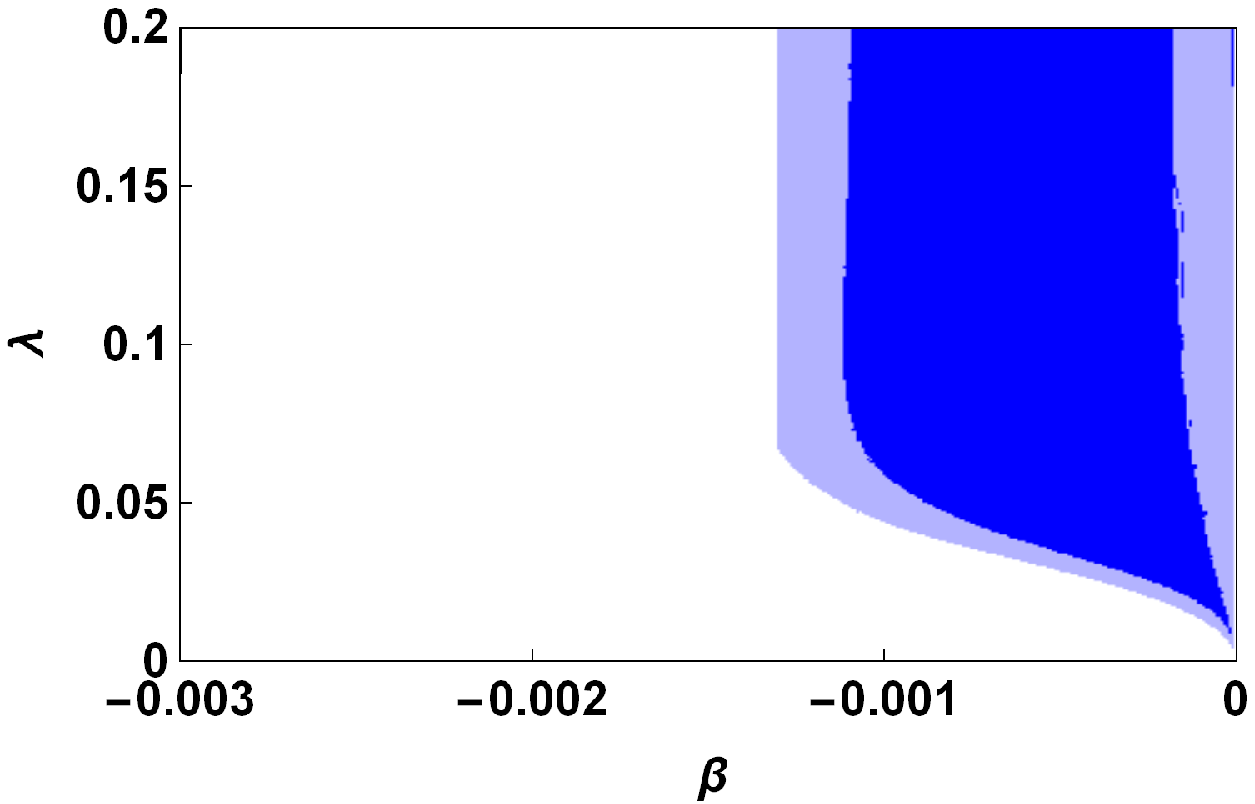}}
  \caption{The valid range of the constants $\beta$ and $\lambda$ to have a consistent result with data for the scalar spectral index and tensor-to-scalar ratio; where in a) $\varepsilon=+1$ and b) $\varepsilon=-1$.}\label{BetaLambda04}
\end{figure}

\begin{figure}[h]
\centering
\includegraphics[width=7cm]{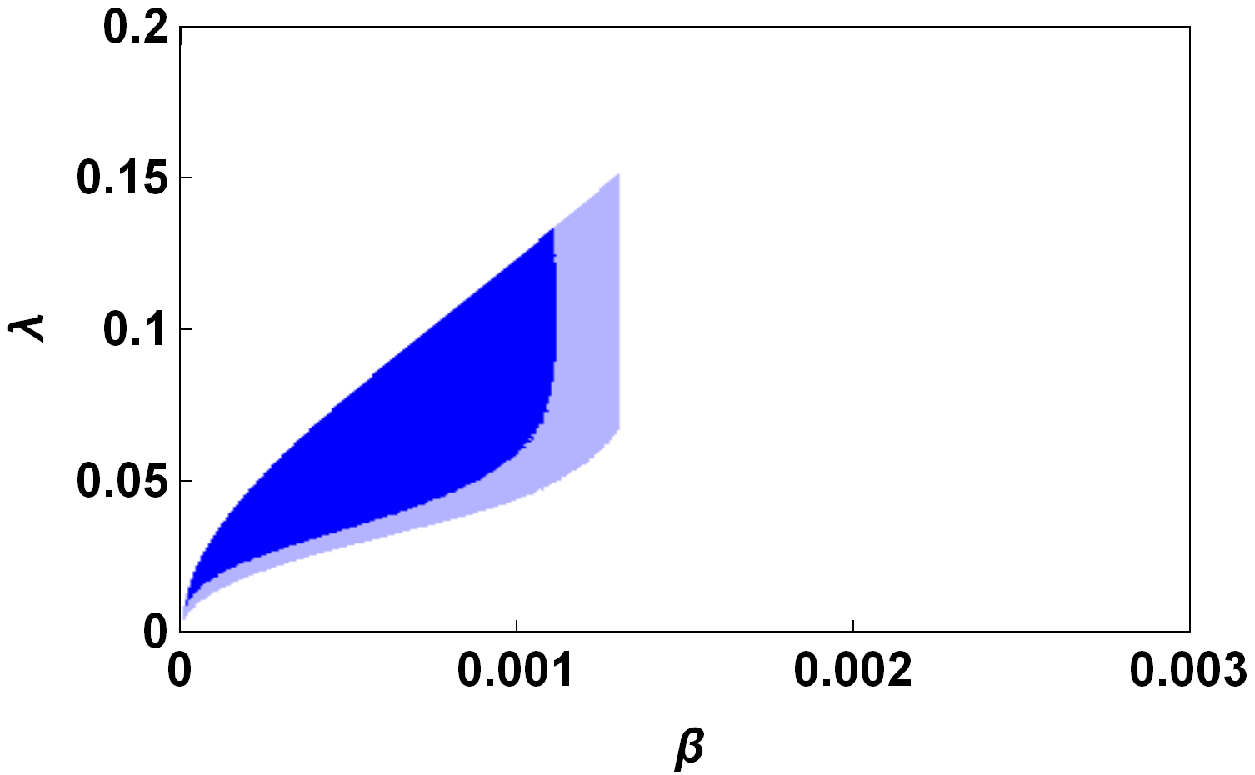}
\caption{Parametric space of $\beta$ and $\lambda$ to give the acceptable values for $n_s$ and $r$, and sound speed $0.1<c_s<1$}\label{csbetalambda04}
\end{figure}

As a final case, the function $f(\phi)$ is taken as a hyperbolic function, $f(\phi)=f_0 \cosh(\lambda \phi)$. Plugging it into Eq.(\ref{hubble}), the Hubble parameter is extracted as a function of scalar field as
\begin{equation}\label{hubble04}
H(\phi) = {\varepsilon \lambda \over 2\beta \sqrt{f_0}} \; {\sinh(\lambda \phi) \over \cosh^{3/2}(\lambda \phi)}.
\end{equation}
Using above result and Eq.(\ref{gamma-urr}), the slow-roll parameter $\epsilon$ is read as a function of scalar field as
\begin{equation}\label{epsilon04}
\epsilon = 2\varepsilon \beta \left( \coth^2(\lambda \phi) - {3 \over 2} \right),
\end{equation}
which make it possible to have a graceful exit from inflation. Fig.\ref{figepsilon04} shows the behavior of $\epsilon$ versus scalar field for different choices of $\beta$ which clearly portrays that it could reaches unity and ends the positive accelerated expansion phase of inflation. \\
Let us consider the situation in an analytical viewpoint. Inflation ends when the slow-roll parameters $\epsilon$ reaches unity. Thus the scalar field at the end of inflation is read as
\begin{equation}\label{phiend04}
\cosh(\lambda \phi_e) = \left[ \; \varepsilon + 3\beta  \over \varepsilon + \beta \; \right]^{1/2}.
\end{equation}
On the other hand, from the equation of number of e-folds, the scalar field at the horizon crossing could be expressed as
\begin{equation}\label{phistar04}
\cosh(\lambda \phi_\star) = \cosh(\lambda \phi_e) \; \exp\left[ {2\varepsilon \beta N} \right]
\end{equation}
As demonstrated in Fig.\ref{figepsilon04}, it is clearly seen that the slow-roll parameter $\epsilon$ reaches unity by decreasing of the scalar field, indicating the presence of an end for the inflation. Also, unlike the previous case the scalar field could be positive which might be more favorable case.\\
\noindent From Eq.(\ref{s}), the slow-roll parameter $s$ which appears in the scalar spectral index is obtained in terms of scalar field as
\begin{equation}\label{s04}
s = 2 \varepsilon \beta \left[ \; { \cosh^2(\lambda \phi) - 9  \over  3 - \cosh^2(\lambda \phi) } + {1 \over 2} \; \right].
\end{equation}
Substituting $\phi_\star$ in Eqs.(\ref{epsilon04}) and (\ref{s04}), and using Eqs.(\ref{ns}) and (\ref{r}), the scalar spectral index and tensor-to-scalar ratio are derived at the time of horizon crossing only in terms of the constants $\beta$ and $\lambda$. Using the Planck data, a parametric space of $\beta$ and $\lambda$ could be depicted, as Fig.\ref{BetaLambda04}, for which the scalar spectral index and tensor-to-scalar ratio of the model come to an agreement with data.
The other free parameter of the model, i.e. $f_0$, is determined through the curvature parameters and by imposing the observational data. From Eq.(\ref{pssuper}), and by applying Eqs.(\ref{hubble04}), (\ref{epsilon04}) and (\ref{phistar04}), $f_0$ is given by
\begin{equation}\label{f04}
f_0 = \left( {2^{\nu - {3 \over 2}} \; \Gamma(\nu) \over \Gamma(3/2)} \right)^2 \; {\lambda^2 \over 4\beta^2 \left( 8 \pi^2 c_s^\star \epsilon^\star \mathcal{P}_s^\star \right)} \; {\sinh^2(\lambda\phi_\star) \over \cosh^3(\lambda\phi_\star)}
\end{equation}
which clearly the dependence on $\beta$ and $\lambda$ is seen. Then, by using Fig.\ref{BetaLambda04} to choose proper values of $\beta$ and $\lambda$, the constant $f_0$ is extracted and in turn, from Eqs.(\ref{potential}) and (\ref{gamma-urr}) the inflation energy scale and the sound speed could be estimated. The results have been implied in Table.\ref{table04} where it could be found out that the energy scale of the inflation at the time of horizon crossing is about $10^{-3}(M_p)$, and the sound speed could be larger than $0.1$ for appropriate choices of $\beta$ and $\lambda$. For every given $\beta$, by increasing $\lambda$, the constant $f_0$ should grow up to result in a consistent amplitude of the scalar perturbations, however, the energy scale of inflation and sound speed reduce. On the other hand, the situation is revered by enhancement of $\beta$ for any given $\lambda$, so that the energy scale of inflation and sound speed increase and the constant $f_0$ decreases.\\
\begin{table}
\centering
  \caption{The estimated values of the constant $f_0$, energy scale of inflation, and sound speed for different values of $\beta$ and $\lambda$. }

\begin{tabular}{lllll}
  \hline
  \hline
  \quad $\beta$ &  $\ \lambda$  & \qquad $f_0$  & \qquad $V^\star$ & $\ c_s^\star$\\
  \hline
  $0.00022$ & $0.05$ & $3.38\times 10^{14}$ & $6.20\times 10^{-12}$ & $0.13$ \\
  $0.00022$ & $0.08$ & $3.35\times 10^{15}$ & $1.51\times 10^{-12}$ & $0.05$ \\
  $0.00048$ & $0.05$ & $8.58\times 10^{12}$ & $1.05\times 10^{-10}$ & $0.39$ \\
  $0.00048$ & $0.08$ & $9.01\times 10^{13}$ & $2.56\times 10^{-11}$ & $0.15$ \\
  $0.00079$ & $0.07$ & $4.01\times 10^{12}$ & $2.46\times 10^{-10}$ & $0.46$ \\
  $0.00079$ & $0.1$  & $2.39\times 10^{13}$ & $8.46\times 10^{-11}$ & $0.22$ \\
  $0.001$   & $0.1$  & $7.12\times 10^{12}$ & $2.12\times 10^{-10}$ & $0.35$ \\
  $0.001$   & $0.15$ & $5.40\times 10^{13}$ & $6.30\times 10^{-11}$ & $0.15$ \\
  \hline
\end{tabular}

\label{table04}
\end{table}
It should be noted that not all the values of $\beta$ and $\lambda$, that have been determined in Fig.\ref{BetaLambda04}, lead to a sound speed larger than $0.1$. To get the desire range for these two constants, one could impose the condition $0.1<c_s<1$ on the parametric space in Fig.\ref{BetaLambda04}. As it is expected, this extra condition shorten the range of the constants $\beta$ and $\lambda$ which is plotted in Fig.\ref{csbetalambda04}. 

\section{New Definition For {\large{$\eta$}}}\label{AnotherEta}
\begin{figure}
  \centering
  \subfigure[$\theta=-1$]{\includegraphics[width=6cm]{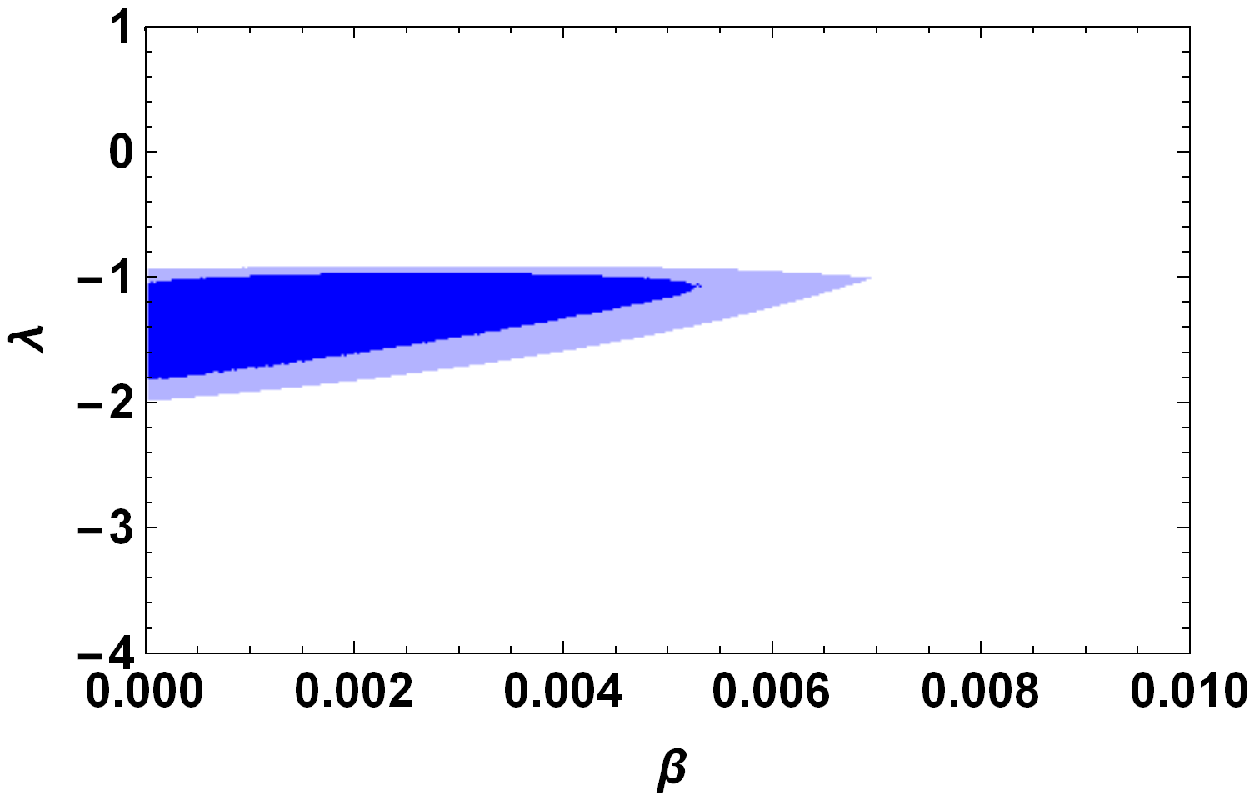}}
  \subfigure[$\theta=+1$]{\includegraphics[width=6cm]{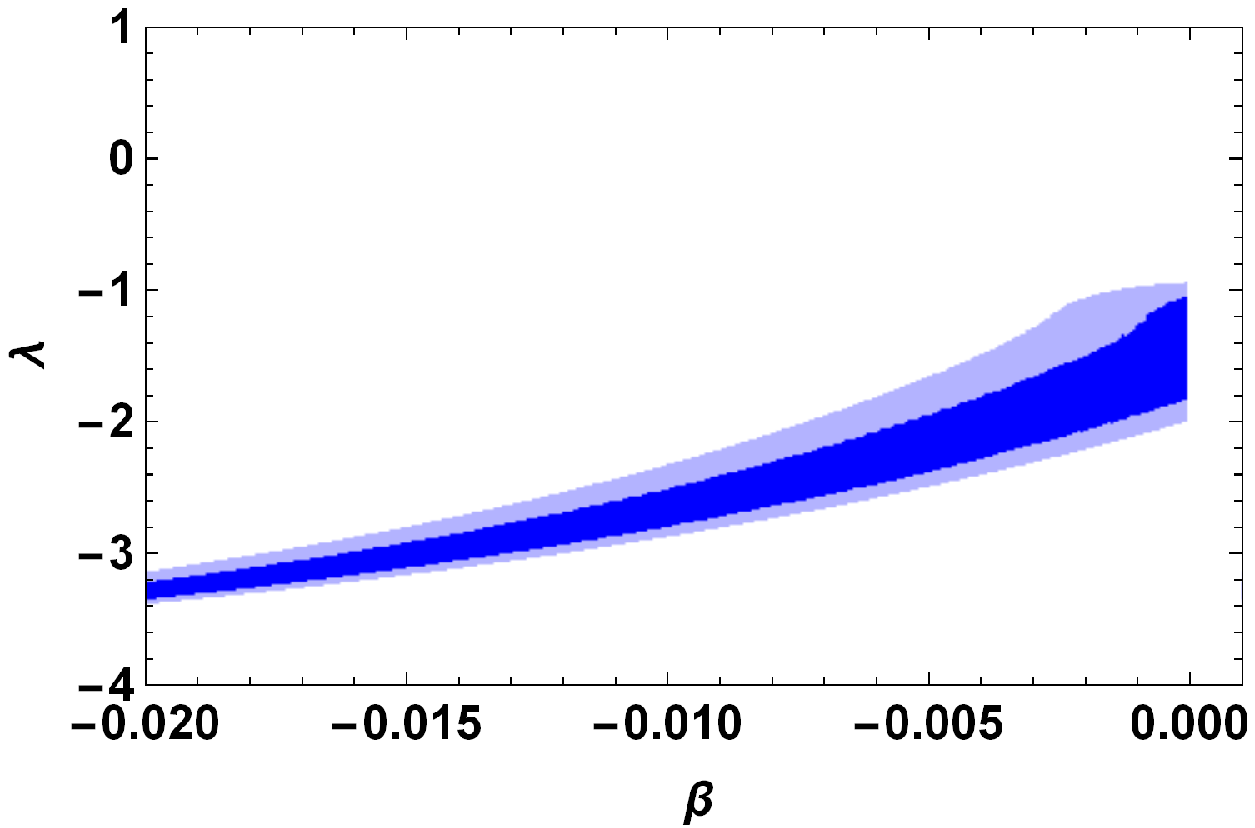}}
  \caption{Parametric space of the free parameters $\beta$ and $n$ for $f_0=5.8\time 10^{6}$ and number of e-fold $N=65$.}\label{bn}
\end{figure}

It is common to define the second slow-roll parameter $\eta$ as the rate of variation of $\dot\phi$ during a Hubble time, namely the definition (\ref{eta}). However, why do we take this definition? A possible answer might be that in the standard model of the (slow-roll) inflation, the equation of motion of scalar field is given as
\begin{equation}\label{EoM}
  \ddot{\phi} + 3H \dot{\phi} - V'(\phi) = 0.
\end{equation}
Then, by imposing this assumption that the $\eta=-\ddot{\phi} / H \dot{\phi}$ is smaller than unity, the term $\ddot\phi$ in the left hand side of the equation could be ignored compared to the second term. Then, smallness of $\eta$ leads one to a simpler equation as $3H\dot\phi = -V'(\phi)$ which is used widely in the inflationary studies. \\
The same viewpoint could be applied in our case. The equation of motion of the DBI scalar field could be written as
\begin{equation}\label{EoMDBI}
{d \over dt}\; (\gamma \dot\phi) + 3 H (\gamma \dot\phi) + {f'(\phi) \over f^2(\phi)} \; \left( 1 - {1 \over \gamma} \right) + V'(\phi) = 0.
\end{equation}
The interesting point is that in ultra-relativistic regime, the term $1/\gamma$ is ignored compared to one, and the remain last two term could be assumed as an effective potential. Then, there is the same equation as Eq.(\ref{EoM}) only we have $(\gamma \dot{\phi})$ instead of $\dot\phi$. It motivates one to define the following expression for the second slow-roll parameter in DBI scalar field model
\begin{equation}\label{neweta}
\eta = {(\gamma \dot{\phi})^{\cdot} \over H (\gamma \dot{\phi})} \; .
\end{equation}
From Eq.(\ref{friedmann}) and (\ref{rhoandp}) we have $\gamma \dot\phi = -2H'(\phi)$. Then, by using Eq.(\ref{phidot}) the differential equation for the Hubble is obtained as
\begin{equation}\label{newHubbleDe}
H''(\phi) - {\varepsilon \beta \over 2}\; H \; \sqrt{ 1 + 4f(\phi) H'^2(\phi) } \; = 0.
\end{equation}
which is still difficult to solve it analytically. In the ultra-relativistic regime, it reduces to
\begin{equation}\label{newHubbleDeff}
H''(\phi) - \varepsilon \beta \; H \; H' \; \sqrt{ f(\phi) } \; = 0.
\end{equation}
Both $f(\phi)$ and $H(\phi)$ was assumed to be a function of scalar field. A way to solve above differential equation is to introduce $f(\phi)$ as a function of the Hubble parameter. Introducing $f(\phi) = f_0 H^n(\phi)$ one arrives at
\begin{equation}\label{Hdphi}
H'(\phi) = {2\varepsilon \beta \sqrt{f_0} \over n+4} \; \left( H^{n+4 \over 2} -c_0 \right) .
\end{equation}
where $c_0$ is a constant of integration. Taking integration, we have
\begin{equation}\label{Hphi}
H \; {}_2F_1\left[ 1 , {2 \over n+4} , 1+{2 \over n+4} , {H^{n+4 \over 2} \over c_0} \right]
= {2 \varepsilon c_0 \beta \sqrt{f_0} \over n+4} \; (\phi + \phi_0)
\end{equation}
in which $\phi_0$ is anther constant of integration. \\
In the rest of this section instead of expressing in terms of the scalar field, we work with the Hubble parameter which brings more convenience. Doing so, the first slow-roll parameter is read as
\begin{equation}\label{newEpsilon}
\epsilon = - {\dot{H} \over H^2} = - {H' \; \dot{\phi} \over H^2} = {2 \varepsilon \beta \over n+4} \;
{H^{n+4 \over 2} - c_0 \over H^{n+4 \over 2}} .
\end{equation}
The inflation ends as $\epsilon = 1$ and the Hubble parameter reaches
\begin{equation}\label{Hend}
H_e^{n+4 \over 2} = {2 \beta c_0 \over 2\beta - \varepsilon (n+4)}.
\end{equation}
The Hubble parameter at the time of the horizon crossing is obtained through the definition of number of e-fold so that
\begin{eqnarray}\label{new-efold}
N & = & \int_{\phi_\star}^{\phi_e} {H \over \dot{\phi}} \; d\phi = \int_{H_\star}^{H_e} {H \over \dot{\phi} \; H'} \; dH \\
 & = & {1 \over \varepsilon \beta} \; \ln\left( {H_\star^{n+4 \over 2} - c_0 \over H_e^{n+4 \over 2} - c_0} \right) .
\end{eqnarray}
and it is found out that the Hubble parameter at the time of horizon crossing is
\begin{equation}\label{Hstar}
H_\star^{n+4 \over 2}  = c_0 \left[ {\varepsilon (n+4) \over 2\beta - \varepsilon (n+4)} \; e^{\varepsilon \beta N} + 1 \right].
\end{equation}
The debates of perturbation is same as Sec.III, and the only difference occurs in the expression of the parameter $\nu$ due to the new definition of $\eta$. The new $\nu$ is given as
\begin{equation}\label{newnu}
\nu^2 = {9 \over 4} + 6\epsilon + 3\eta - {3 \over 2} \; s + 9 \epsilon \eta - \eta s + \eta^2 + 2 \epsilon \eta^2 .
\end{equation}
and all other perturbation parameters have the same behavior as Sec.III. \\
The method for constraining the free parameters of the model is almost same as before. Through the amplitude of scalar perturbations one could the constant $c_0$ as
\begin{equation}\label{c0}
c0^2 = {8\pi^2 \mathcal{P}_s \over 2f_0} \; \left( \Gamma(3/2) \over 2^{\nu-{3\over 2}} \Gamma(\nu) \right)^2 \;
\left[ 2\beta - \varepsilon (n+4)  \over  2\beta + \varepsilon (n+4) \Big( e^{\varepsilon \beta N} - 1 \Big) \right]^2.
\end{equation}
Computing the scalar spectral index and tensor-to-scalar ratio it is realized that they depends on the parameter $\beta$, $n$ and $c_0$. Then, using Eq.(\ref{c0}), and Planck $r-n_s$ diagram, one could find the acceptable values of the free parameter of the free model which is presented in Fig.\ref{bn}.
It is clear that to have a consistency with data, the parameter $n$ should be negative, and the second slow-roll parameter is of order $10^{-3}$ and also it could be of order $10^{-2}$ for $n>-3$ and positive $\theta$.

\begin{figure}
  \centering
  \includegraphics[width=6cm]{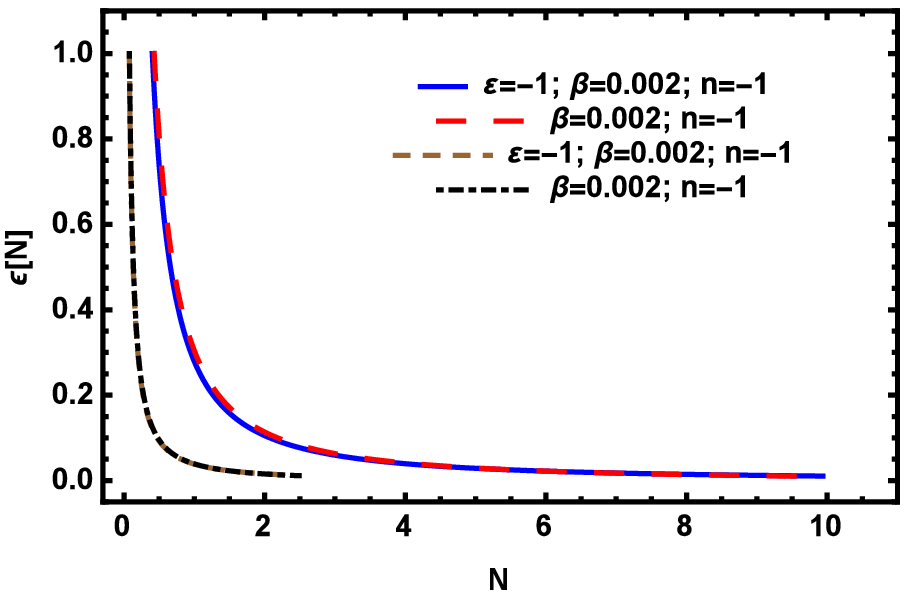}
  \caption{Behavior of the parameter $\epsilon$ during inflation for various values of the free parameters of the model.}\label{newepsilon}
\end{figure}

By finding the Hubble parameter at the beginning and end of inflation and determining the free parameters of the model, one could find out about the behavior of the first slow-roll parameter $\epsilon$ during inflation. Fig.\ref{newepsilon} illustrates the parameter during inflation for various values of $\beta$ and $n$, and all of them state that $\epsilon$ is small at the beginning of inflation and it grows by passing time and decreasing the Hubble parameter, and finally reaches one; where inflation ends.
To get more insight of the case, the following table is presented which give information about the sound speed and energy scale of inflation.
\begin{table}
\centering
  \caption{The estimated values of the constant $c_0$, energy scale of inflation, and sound speed for different values of $\beta$ and $n$. }

\begin{tabular}{lllll}
  \hline
  \hline
  \quad $\beta$ &  $\ n$  & \qquad $c_0$  & \qquad $V^\star$ & $\ c_s^\star$\\
  \hline
  $0.002$ & $-1.0$ & $9.70\times 10^{-7}$ & $2.43\times 10^{-8}$ & $0.37$ \\
  $0.003$ & $-1.1$ & $6.70\times 10^{-7}$ & $6.96\times 10^{-9}$ & $0.17$ \\
  $0.001$ & $-1.2$ & $1.88\times 10^{-6}$ & $1.81\times 10^{-8}$ & $0.073$ \\
  $0.004$ & $-1.2$ & $5.18\times 10^{-7}$ & $2.15\times 10^{-9}$ & $0.081$ \\
  $0.001$ & $-1.5$ & $1.88\times 10^{-6}$ & $1.86\times 10^{-9}$ & $0.004$ \\
  $0.002$ & $-1.5$ & $9.70\times 10^{-7}$ & $5.80\times 10^{-10}$ & $0.004$ \\
  \hline
  $-0.0012$   & $-1.5$  & $1.58\times 10^{-6}$ & $1.37\times 10^{-9}$ & $0.004$ \\
  $-0.0048$   & $-2$  & $4.42\times 10^{-7}$ & $3.08\times 10^{-13}$ & $6.69 \times 10^{-6}$ \\
  $-0.0088$   & $-2.5$  & $2.71\times 10^{-7}$ & $1.93\times 10^{-18}$ & $1.41 \times 10^{-10}$ \\
  $-0.015$   & $-3$ & $1.89\times 10^{-7}$ & $6.89\times 10^{-29}$ & $7.15 \times 10^{-20}$ \\
  \hline
\end{tabular}

\label{tableNew}
\end{table}
The sound speed could get various values, but the important point is that for $\theta=+1$, the proper values of $\beta$ and $n$ which come from Fig.\ref{bn}, could lead to a very low sound speed and also a very low energy scale.

\section{Attractor Behavior}\label{Attractor}
\begin{figure}
\centering
\subfigure[]{\includegraphics[width=7cm]{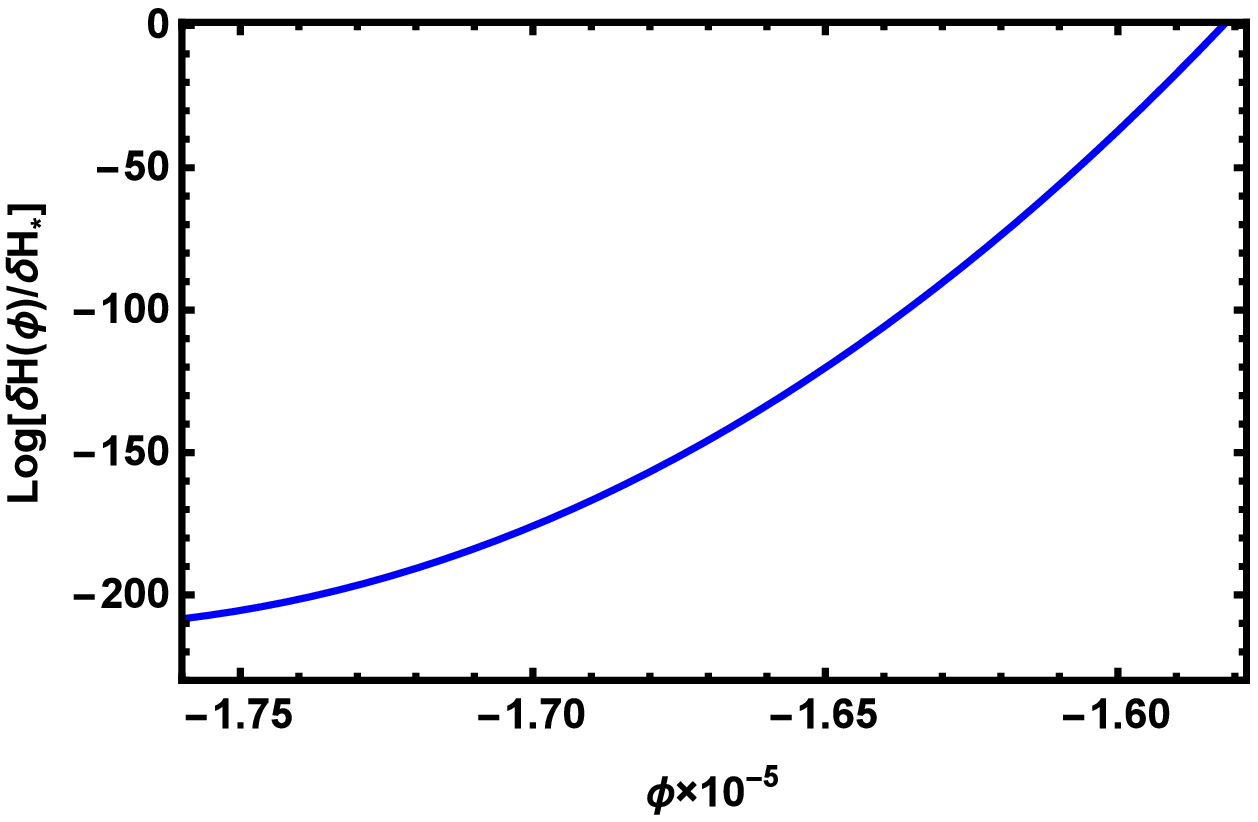}}
\subfigure[]{\includegraphics[width=7cm]{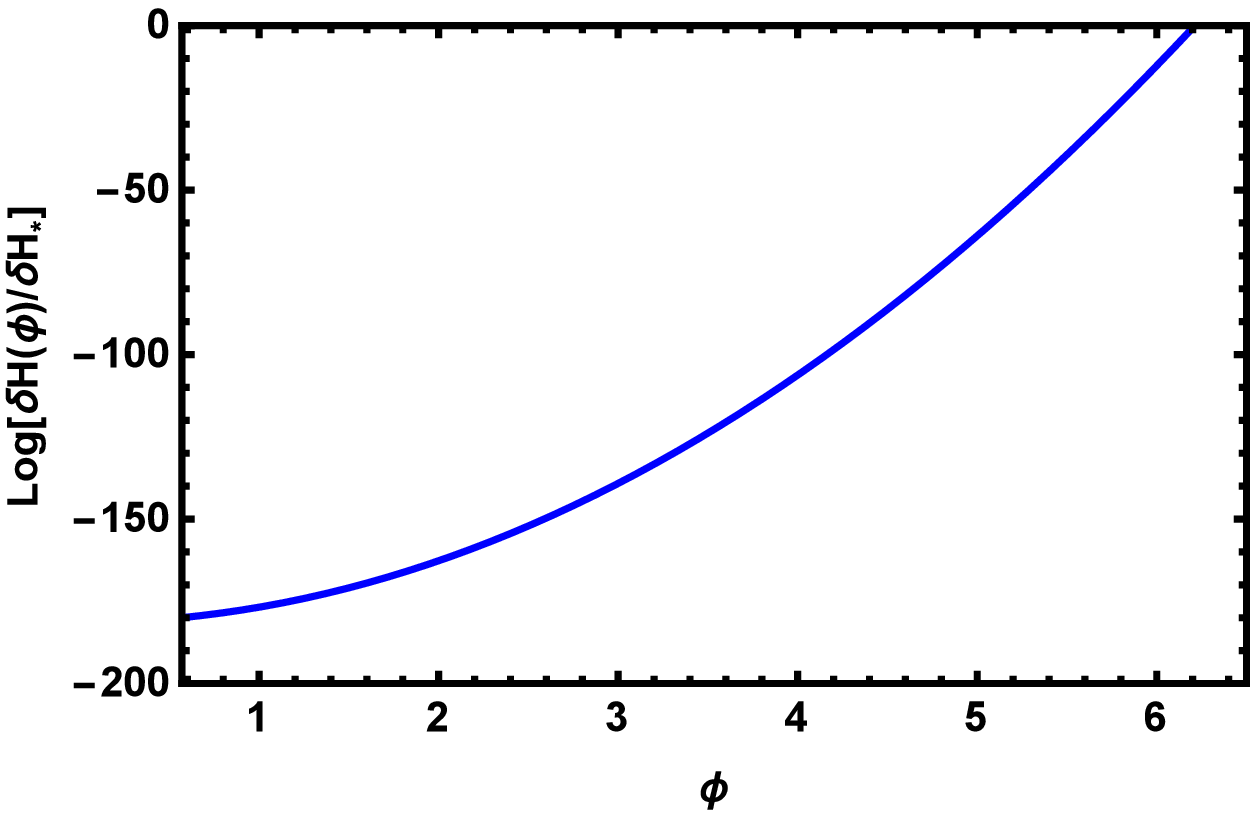}}
\caption{The logarithm of the perturbation parameter $\delta H(\phi)/\delta H_\star$ in terms of the scalar field during the inflation. By passing time, the scalar field decreases and the logarithm tends to bigger negative values which implies that the perturbation parameter rapidly decrease by approaching to the end of inflation.\label{figattractor}}
\end{figure}

Attractor behavior is a feature that should be considered for the inflationary solutions. Hamilton-Jacobi approach provides a suitable way to consider the attractor behavior of the solution which is introduced in \citep{Lyth}. Following this method, we assume a homogenous perturbation for the Hubble parameter, i.e $\delta H=H-H_0$, and substitute this in the Hamilton-Jacobi equation (\ref{potential}). Computing the equation up to the first order of the perturbation results in the following equation
\begin{equation}\label{attractor}
\delta H(\phi) = \delta H_\star \exp\left( 3 \int_{\phi_\star}^{\phi} \; H(\phi) \sqrt{f(\phi)} \; d\phi \right).
\end{equation}
in which $H_\star$ is the perturbation at the initial time. In order to satisfy the attractor behavior of the solution, the perturbation $\delta H(\phi)$ reduces during the inflation. For the third and forth cases, there are
\begin{eqnarray}\label{attractor3-4}
\delta H(\phi) & = & \delta H_\star \exp \left( {\varepsilon \over 2\beta} \; \int_{\phi_\star}^{\phi} {1 + \lambda \phi \over \phi} \; d\phi \right), \nonumber \\
\delta H(\phi) & = & \delta H_\star \exp \left( {\varepsilon \lambda \over 2\beta} \; \int_{\phi_\star}^{\phi} {\sinh(\lambda \phi) \over \cosh(\lambda \phi) } \; d\phi \right).
\end{eqnarray}
Fig.\ref{figattractor} illustrates the behavior of the perturbation $\delta H(\phi)$ versus the scalar field during the inflation. Since the perturbation rapidly vanishes the logarithm of the perturbation is depicted which states that at the initial time where $\delta H(\phi)/\delta H_\star=1$ the logarithm is zero and it is decreases and tends to the negative values as time passes and inflation ends.

\section{Conclusion}\label{conslusion}
The constant-roll inflationary scenario in the framework of Einstein gravity was studied in which the inflaton was assumed to be a DBI scalar field. By acquiring the main dynamical equations of the model and applying Hamilton-Jacobi formalism, where the scalar field plays the role of time and the Hubble parameter is assumed to be a function of scalar field, a non-linear differential equation for the Hubble parameter was obtained. Gaining an analytical solution for this differential equation encountered with difficulties, then the work was restricted to the ultra-relativistic regime where the Lorentz factor $\gamma$ is large. In this regime, the time derivative of scalar field was realized to be negative, consequently a decreasing behavior for the scalar field was expected during the inflationary times. \\
Comparing the results with the observational data was the main purpose of the presented work. In this regard, considering the cosmological perturbations of the model was required which was investigated in Sec.\ref{SecIII}. Since the second slow-roll parameter is assumed to be a constant, which might not be small, some extra terms in the expression of the amplitude of scalar perturbations appear that in turn affect the scalar spectral index. On the other side, since perturbations of the energy-momentum tensor has no contribution in tensor perturbations, and because of this fact that only the slow-roll parameter $\epsilon$ contributes in the expression of the tensor perturbations, there was no modification in terms of the slow-roll parameters in the amplitude of tensor perturbations. Deriving the main perturbation parameters for the model, the consistency of the model predictions with observational data was performed in Sec.\ref{SecIV} for specific choices of $f(\phi)$ function.  \\
Attributing a power-law and exponential functions to $f(\phi)$ leads to a constant slow-roll parameter $\epsilon$. This result could not give a graceful exit from inflation, although might leads to an accelerated expansion phase by properly selecting the free parameters of the model. As the third case, $f(\phi)$ function was taken as a product of linear and exponential function of scalar field. This choice results in a varying slow-roll parameter $\epsilon$ which also could give a graceful exit, i.e. $\epsilon=1$. However, to have decreasing behavior of the scalar field during inflation, as it is required by Eq.(\ref{gamma-urr}), the scalar field during inflation should be negative; plotted in Fig.\ref{figepsilon03}.  Computing the scalar spectral index and the tensor-to-scalar ratio determined that they only depend on the constant $\beta$ and $\lambda$. Then, using the Planck $r-n_s$ diagram we could obtained a range for these two free constant of the model which put the model prediction in great consistency with the data. It shows that the second slow-roll parameter is small and of order $10^{-2}$ or $10^{-1}$, however for the latest order the energy scale of the inflation and the sound speed is very low. On the other hand for the former order of $\beta$, the energy scale of the inflation is around $10^{-3} M_p$ and the sound speed could be larger than $0.1$. Also, the amplitude of scalar perturbation was used to determined the third free constant of the model, i.e. $f_0$. \\
As the last typical example, the function $f(\phi)$ was taken as a hyperbolic function of the scalar field. This choice leads to a varying $\epsilon$ which could produce a graceful exit from inflation after enough expansion. In contrast to the third case, to satisfy decreasing behavior of scalar field during inflation, the scalar field is not required to be negative. The calculation indicates that at the time of horizon crossing, the scalar spectral index and tensor-to-scalar ratio only depends on the two constants of the model, i.e. $\beta$ and $\lambda$. Using the $r-n_s$ diagram, we could determine the range of these two constants that leads to a consistent result with data. It states that the second slow-roll parameter could be of order $10^{-3}$. The other constant of the model was constrained through the amplitude of the scalar perturbation. Using these outcomes, the energy scale of inflation was found to be of order $10^{-3} M_p$ and the sound speed was found to be larger than $0.1$.  \\
Comparing the equation of motion of DBI scalar field with the corresponding in standard model comes to an interesting idea. Here it is $(\gamma \dot\phi)$ that plays the same role that $\dot\phi$ has in standard model of scalar field. Then a new definition was given to $\eta ={(\gamma \dot\phi)^{\cdot} \over H (\gamma \dot\phi)}$. It resulted in different solution and changes in perturbation parameters. Following the same procedure we could determine the range for the constants of the model which put the result in good consistency with the data.  \\
The attractor behavior of the solutions was investigated for the last two cases. Assuming a homogeneous perturbation for the Hubble parameter, it was determined that by passing time the perturbation decreases implying that the feature could be satisfied properly.

\section{Acknowledgments} 
AM would like to thanks the ”Ministry of Science, Research and Technology” of Iran for financial support, and Prof. R. Casadio for hospitality during his visit. The work of T. G. has been supported financially by ”Vice Chancellorship of Research and Technology, University of Kurdistan” under research Project No. 98/11/2724.






\end{document}